\documentclass[10pt,journal,compsoc]{IEEEtran}

\ifCLASSOPTIONcompsoc
  \usepackage[nocompress]{cite}
\else
  \usepackage{cite}
\fi

\usepackage{multirow} 

\usepackage{subfigure}
\usepackage{pifont}
\usepackage{xcolor}
\usepackage{bm}
\usepackage{url}
\usepackage{hyperref}
\hypersetup{
    colorlinks=true,
    linkcolor=blue,
    filecolor=magenta,      
    urlcolor=cyan,
}

\usepackage{makecell}

\usepackage{algorithm}
\usepackage{algorithmic}
\usepackage{setspace}

\usepackage{pifont}
\newcommand{\circled}[1]{\textcircled{\raisebox{-.9pt}{#1}}}
\usepackage{ragged2e}
\usepackage{graphicx} 
\usepackage{amsmath} 

\newtheorem{definition}{Definition}

\newcommand{\toolname}{\textsc{Esale}}

\usepackage[normalem]{ulem}

\newcommand{\revise}[1]{{\color{black}{#1}}}
\newcommand{\delete}[1]{}

\newcommand{\deletion}[1]{}

\newcommand{\revision}[1]{{\color{black}{#1}}}

\newcommand{\deleted}[1]{}

\newcommand{\revised}[1]{{\color{black}{#1}}}

\begin{document}

\title{{\toolname}:
\underline{E}nhancing
Code-\underline{S}ummary \underline{A}lignment \underline{Le}arning for Source Code Summarization}

\author{Chunrong~Fang, Weisong~Sun*, Yuchen~Chen, Xiao~Chen, Zhao Wei, Quanjun~Zhang, Yudu~You, Bin~Luo, Yang Liu, Zhenyu~Chen

\IEEEcompsocitemizethanks{\IEEEcompsocthanksitem Chunrong Fang, Yuchen Chen, Xiao Chen, Quanjun Zhang, Yudu You, Bin Luo and Zhenyu Chen are with the State Key Laboratory for Novel Software Technology, Nanjing University, Nanjing, China and also with the Software Institute, Nanjing University, Nanjing, Jiangsu 210008, China.
E-mail: fangchunrong@nju.edu.cn, yuc.chen@outlook.com, shawnchan@smail.nju.edu.cn, quanjun.zhang@smail.nju.edu.cn, fzuyyd@163.com, luobin@nju.edu.cn, zychen@nju.edu.cn.

\IEEEcompsocthanksitem Weisong Sun and Yang Liu are with the College of Computing and Data Science, Nanyang Technological University, Singapore.
E-mail: weisong.sun@ntu.edu.sg, yangliu@ntu.edu.sg.

\IEEEcompsocthanksitem Zhao Wei is with Tencent Inc., China. Email: zachwei@tencent.com.

\IEEEcompsocthanksitem *Weisong Sun is corresponding author.}

\thanks{Manuscript received xxx xxx, 2023; revised xxx xxx, 2024.}}

\markboth{Transaction on Software Engineering,~Vol.~xxx, No.~xxx, xxx~2024}%
{Shell \MakeLowercase{\textit{et al.}}: Bare Demo of IEEEtran.cls for Computer Society Journals}


\IEEEtitleabstractindextext{%
\begin{abstract}
\justifying
(Source) code summarization aims to automatically generate succinct natural language summaries for given code snippets. Such summaries play a significant role in \deletion{facilitating}\revision{promoting} developers to understand and maintain code. 
Inspired by neural machine translation, deep learning-based code summarization techniques widely adopt an encoder-decoder framework, where the encoder transforms given code snippets into context vectors, and the decoder decodes context vectors into summaries.
Recently, large-scale pre-trained models for source code (e.g., CodeBERT and UniXcoder) are equipped with encoders capable of producing general context vectors and have achieved substantial improvements on the code summarization task. 
However, although they are usually trained mainly on code-focused tasks and can capture general code features, they still fall short in capturing specific features that need to be summarized. In a nutshell, they fail to learn the alignment between code snippets and summaries (code-summary alignment for short).

In this paper, we propose a novel approach to improve code summarization based on summary-focused tasks. Specifically, we exploit a multi-task learning paradigm to train the encoder on three summary-focused tasks to enhance its ability to learn code-summary alignment, including unidirectional language modeling (ULM), masked language modeling (MLM), and action word prediction (AWP). Unlike pre-trained models that mainly predict masked tokens in code snippets, we design ULM and MLM to predict masked words in summaries. Intuitively, predicting words based on given code snippets would help learn the code-summary alignment. In addition, existing work shows that AWP affects the prediction of the entire summary. Therefore, we further introduce the domain-specific task AWP to enhance the ability of the encoder to learn the alignment between action words and code snippets. 
We evaluate the effectiveness of our approach, called {\toolname}, by conducting extensive experiments on \delete{two widely used}\revise{four} datasets\revise{, including two widely used datasets JCSD and PCSD, a cross-project Java dataset CPJD, and a multilingual language dataset CodeSearchNet}. Experimental results show that {\toolname} significantly outperforms state-of-the-art baselines in \deletion{terms of }all three widely used metrics, including BLEU, METEOR, and \delete{ROUGH-L}\revise{ROUGE-L}. \deletion{In addition}\revision{Moreover}, the human evaluation proves that the summaries generated by {\toolname} are more informative and closer to the ground-truth summaries.
\end{abstract}

}

\maketitle

\IEEEdisplaynontitleabstractindextext

%
\IEEEpeerreviewmaketitle

\graphicspath{{figures/}} 

\IEEEraisesectionheading{\section{Introduction}\label{sec:introduction}}
\IEEEPARstart{C}{ode} comments play a key role in facilitating code comprehension~\cite{1981-Comments-on-Program-Comprehension, 1988-Program-Readability, 2018-Measuring-Program-Comprehension, 2020-Code-to-Comment-Translation} and software maintenance~\cite{1993-Maintenance-Productivity, 2005-Documentation-Essential-Software-Maintenance, 2006-Code-Comments-in-PostgreSQL, 2020-CPC, 2021-Why-My-Code-Summarization-Not-Work}. For example, prior works~\cite{1995-Procedures-and-Comments, 1988-Program-Readability, 1981-Comments-on-Program-Comprehension} show that code comments can help improve code readability. Commenting code has been recognized as a good programming practice~\cite{2005-Documentation-Essential-Software-Maintenance, 2020-CPC}. However, writing high-quality code comments is a labor-intensive and time-consuming task~\cite{2005-Documentation-Essential-Software-Maintenance, 2005-Survey-of-Documentation-Practice}. As a result, good comments are often absent, unmatched, and outdated during the software evolution~\cite{2018-TL-CodeSum}.\deletion{ The lack of high-quality code comments is a common problem in the software industry~\cite{2018-TL-CodeSum}.} (Source) code summarization is an active research field~\cite{2010-Automated-Text-Summarization-Summarizing-Code, 2013-Evaluating-Source-Code-Summarization, 2018-Study-of-StaQC-Code-Summarization, 2019-Automatic-Code-Summarization, 2019-Datasets-for-Code-Summarization, 2019-Eye-Movements-in-Code-Summarization, 2020-Human-Study-Code-Summarization, 2020-Code-to-Comment-Translation, 2021-Action-Word-Prediction-for-Code-Summarization, 2022-Evaluation-Neural-Code-Summarization, 2021-Why-My-Code-Summarization-Not-Work, 2021-Reassessing-Metrics-for-Code-Summarization, 2021-Adversarial-Robustness-Deep-Comment-Generation}, which aims at designing advanced techniques to support automatic generation of code comments (also called summaries). Given a code snippet (a method or function) by the developer, code summarization can generate summaries describing the functionality of the code snippet. 

Existing code summarization techniques can mainly be categorized into \textit{keywords-based methods}\deletion{~\cite{2010-Program-Comprehension-with-Code-Summarization, 2010-Towards-Generating-Summary-Java-Methods, 2010-Automated-Text-Summarization-Summarizing-Code, 2013-Automatic-Generation-Summaries-for-Java-Classes, 2014-Code-Summarization-of-Method-Context}}, \textit{retrieval-based methods}\deletion{~\cite{2015-CloCom, 2020-Retrieval-based-Neural-Code-Summarization, 2020-R2Com, 2021-EditSum}}, and \textit{deep learning-based methods}\deletion{~\cite{2018-DeepCom, 2019-Code-Summarization-with-Extended-Tree-LSTM, 2020-Transformer-based-Approach-for-Code-Summarization, 2021-SiT}}. \textit{Keywords-based methods} extract critical terms from code snippets to constitute their summaries\revision{~\cite{2010-Towards-Generating-Summary-Java-Methods, 2010-Automated-Text-Summarization-Summarizing-Code}}. \deletion{For example, the work~\cite{2010-Program-Comprehension-with-Code-Summarization} adopts the Latent Semantic Analysis (LSA) techniques~\cite{2009-Update-Summarization-based-LSA} to determine the importance of every term in the code snippet and then selects the top 5 important terms to compose the summary. }Such methods may fail to generate accurate \deletion{comments}\revision{summaries} if the source code contains poorly named identifiers or method names~\cite{2015-CloCom, 2018-DeepCom}. \textit{Retrieval-based methods} first leverage code clone detection techniques to retrieve similar code snippets and then use their corresponding comments to summarize other code snippets. Similar code snippets can be retrieved from existing open-source platforms (e.g., GitHub~\cite{2015-CloCom}) or software Q\&A sites (e.g., Stack Overflow)~\cite{2013-AutoComment}. Such methods rely on whether similar code snippets can be retrieved~\cite{2021-EditSum} and how similar the \revision{code }snippets are~\cite{2018-DeepCom}. In addition, code snippets may contain some information inconsistent with the content in comments of their similar code snippets~\cite{2020-R2Com}, making retrieval-based methods ineffective in many cases. \textit{Deep learning-based methods} leverage powerful generative models trained on a large-scale code-comment corpus to translate code snippets in programming languages into summaries in natural language\deletion{~\cite{2020-Code-to-Comment-Translation, 2022-Evaluation-Neural-Code-Summarization, 2019-Ast-attendgru, 2018-Code2seq, 2018-DeepCom, 2016-CODE-NN}}\revision{~\cite{2016-CODE-NN, 2018-DeepCom}}. Such methods can model the semantic mapping relations between code snippets and summaries and can generate high-quality summaries~\cite{2016-CODE-NN}. 

Recently, with the success of the pre-training and fine-tuning paradigm in the field of \deletion{NLP}\revision{ natural language processing (NLP)} (e.g., BERT~\cite{2019-BERT} and T5~\cite{2020-T5}), many works in software engineering\revision{ (SE)} have introduced this paradigm to boost further code-related tasks, including code summarization (e.g., CodeBERT~\cite{2020-CodeBERT}, CodeT5~\cite{2021-CodeT5}, and UniXcoder~\cite{2022-UniXcoder}). In practice, these works \deletion{first}\revision{typically} pre-train a model with general language modeling tasks, such as masked language modeling (MLM)~\cite{2019-BERT} and unidirectional language modeling (ULM)~\cite{2018-Unidirectional-Language-Modeling}\deletion{. Then, they fine-tune the pre-trained models }\revision{, followed by fine-tuning }on \revision{code summarization tasks.}\deletion{downstream tasks, such as code clone~\cite{2021-CodeT5, 2022-UniXcoder}, code search~\cite{2020-CodeBERT, 2022-UniXcoder}, and code summarization~\cite{2020-CodeBERT, 2021-CodeT5, 2022-UniXcoder}. }
In \textit{\deletion{Deep}\revision{deep} learning-based methods}\deletion{ (including pre-trained models-based methods)}, code summarization is widely considered as the neural machine translation (NMT) task where the source text is the code snippet in programming language and the target text is the summary in natural language~\cite{2019-Ast-attendgru, 2019-Code-Summarization-with-Extended-Tree-LSTM, 2020-Hybrid-DeepCom, 2020-Code-to-Comment-Translation}. \deletion{Therefore, the goal of training a code summarization model is to learn the mapping between code snippets and summaries. }Inspired by NMT, code summarization models widely adopt the encoder-decoder framework. The encoder is responsible for transforming the code snippet\deletion{ given by the developer} into a context vector. The decoder is responsible for decoding the context vector into a natural language summary. Intuitively, context vectors tell the decoder what content needs to be translated, which indicates that the encoder plays a significant role in a code summarization model. Therefore, to achieve high-quality code summarization, a good encoder should be able to produce context vectors that capture the code features that need to be translated by the decoder.
However, although the advanced pre-trained encoders have achieved significant progress in producing general vector representations (i.e., context vectors) for given code snippets, they are still insufficient in capturing specific code features that need to be translated. These pre-trained encoders are primarily trained with code-focused tasks that teach them to learn the relationship among the tokens of code snippets rather than the relationship between code snippets and summaries.
In other words, these pre-trained encoders are still insufficient in capturing the alignment between code snippets and summaries (i.e., code-summary alignment), detailed in Section~\ref{sec:motivating_example}.

In this paper, we propose a novel approach to improve code summarization. Our approach is built upon large-scale pre-trained code encoders that have been shown to be superior in capturing and representing general code features. To improve the ability of our encoder to learn the code-summary alignment, we exploit the multi-task learning \revision{(MTL) }paradigm to train it. 
In the \deletion{multi-task learning}\revision{MTL} paradigm, multiple tasks are simultaneously learned by a shared model~\cite{2020-Multi-Task-Learning-Survey, 2019-Multi-Task-DNN-for-NLU, 2011-NLP-Almost-from-Scratch}. Such a paradigm can improve data efficiency, reduce overfitting through shared representations, and accelerate learning speed by leveraging auxiliary information. 
\revision{SE researchers have also introduced MTL to address SE tasks. For example, Aung et al.~\cite{2022-Multi-triage} find that both two important tasks, i.e., developer recommendation and issue type classifying involved in the bug triage process, rely on historical bug descriptions and code snippet information. Therefore, they train a multi-triage model to resolve both tasks simultaneously via MTL, and demonstrate this model can reduce task repetition and leverage the correlating information between tasks. MTL has also demonstrated promise in training a pre-trained code representation model (e.g., MulCode~\cite{2021-MulCode} and UniXcoder) and achieved promising efficacy on various downstream SE tasks.} 
In our setting, we perform \deletion{multi-task learning}\revision{MTL} on three summary-focused \revision{pre-training }tasks, including \deletion{unidirectional language modeling (ULM)}\revision{ULM}, \deletion{masked language modeling (MLM)}\revision{MLM}, and action word prediction (AWP)~\cite{2021-Action-Word-Prediction-for-Code-Summarization}. The three tasks are simultaneously learned by a shared encoder. 
ULM and MLM are two general tasks borrowed from the field of \deletion{natural language processing (NLP)}\revision{NLP}. ULM and MLM are important because they can facilitate the language model to capture the relationships of words in the text~\cite{2018-Unidirectional-Language-Modeling, 2019-BERT}. Unlike pre-trained models (e.g., CodeBERT and UniXcoder) that predict masked code tokens based on unmasked parts of code snippets, we design ULM and MLM to predict masked words in summaries based on code snippets. Intuitively, predicting masked words in summaries based on code snippets would help the encoder learn the alignment between masked words and code snippets. In addition, existing work~\cite{2021-Action-Word-Prediction-for-Code-Summarization} shows that \deletion{action word prediction}\revision{AWP} affects the prediction of subsequent words and thus the prediction of the entire summary. Therefore, we further introduce the domain-specific task AWP to enhance the ability of the encoder to learn the alignment between action words and code snippets. In summary, all three summary-focused tasks can enhance the encoder to learn the code-summary alignment such that it can capture the specific code features that need to be summarized. 
\revision{In practice, to reduce the training cost, instead of training the shared encoder from scratch, we initialize the shared encoder with an existing pre-trained encoder, e.g., UniXcoder's encoder.} 
\revise{After obtaining the \deletion{well-trained}\revision{pre-trained} shared encoder, we further train a code summarization model capable of generating a succinct natural language summary for a given code snippet. Specifically, we fine-tune the \deletion{well-trained}\revision{pre-trained} shared encoder on the code summarization task and simultaneously train a decoder.}
\deletion{We evaluate the effectiveness of our approach, called {\toolname}, by conducting extensive experiments on \delete{two widely used datasets (a Java dataset and a Python dataset)}\revise{four datasets, including, two widely used Java and Python datasets (called JCSD and PCSD), a cross-projected Java dataset (called CPJD), and a multilingual dataset (called CSN), detailed in Section~\ref{subsubsec:dataset}}. 
The experimental results show that {\toolname} is significantly better than state-of-the-art baselines in terms of all three widely used metrics, including BLEU, METEOR, and ROUGE-L. In addition, we conduct a human study to evaluate the summaries generated by {\toolname} from four aspects: similarity, naturalness, informativeness, and relevance. And the human evaluation results validate that the summaries generated by {\toolname} are more informative and relevant, and closer to the ground-truth summaries.}

In summary, we make the following contributions.
\begin{itemize}
    \item We propose a novel approach \revision{named {\toolname} }to improve code summarization\deletion{, which uses}\revision{. {\toolname} devises} three summary-focused pre-training tasks (two general tasks ULM and MLM\revision{,} and one domain-specific task AWP) to enhance the encoder to learn the code-summary alignment.
    
    \item We introduce a domain-specific task (i.e., AWP) as one of the important pre-training tasks. We perform an in-depth analysis of the effect of the AWP task, and statistical results show that this task can significantly improve code summarization performance (detailed in Section~\ref{subsubsec:results_of_RQ2}). 
    
    \item We conduct extensive quantitative experiments on \delete{two widely used}\revise{four} datasets\revision{, including two widely used Java and Python datasets (called JCSD and PCSD), a cross-projected Java dataset (called CPJD), and a multilingual dataset (called CSN),} to evaluate \deletion{our approach,}{\toolname}. Experimental results show that {\toolname} significantly outperforms \deletion{state-of-the-art }baselines in terms of all three widely used automatic metrics BLEU, METEOR, and ROUGE-L (detailed in Section~\ref{subsubsec:result_of_RQ1}). The source code of {\toolname} and all the data used in this paper are released and can be downloaded from the website~\cite{2024-ESALE}. 
    
    \item We conduct a qualitative human evaluation to evaluate the summaries generated by {\toolname} and baselines in terms of four aspects: similarity, naturalness, informativeness, and relevance. And statistical results of human scores show that the summaries generated by {\toolname} are more informative and closer to the ground-truth summaries (detailed in Section~\ref{subsubsec:human_evaluation}).
\end{itemize}

The rest of this paper is organized as follows. \deletion{Section~\ref{sec:background} describes the background of four concepts. }Section~\ref{sec:motivating_example} illustrates the motivation of this paper. Section~\ref{sec:methodology} introduces the design of {\toolname}. Section~\ref{sec:evaluation} presents the design of the experiments in detail and gives the details of experiment results and analysis. Section~\ref{sec:case_study} presents a case study. Section~\ref{sec:threats_to_validity} introduces threats to validity. Section~\ref{sec:related_work} discusses the related work. We conclude the paper in Section~\ref{sec:conclusion}.

\section{Motivating Example}
\label{sec:motivating_example}

\begin{figure}[htbp]
  \centering
  \includegraphics[width=\linewidth]{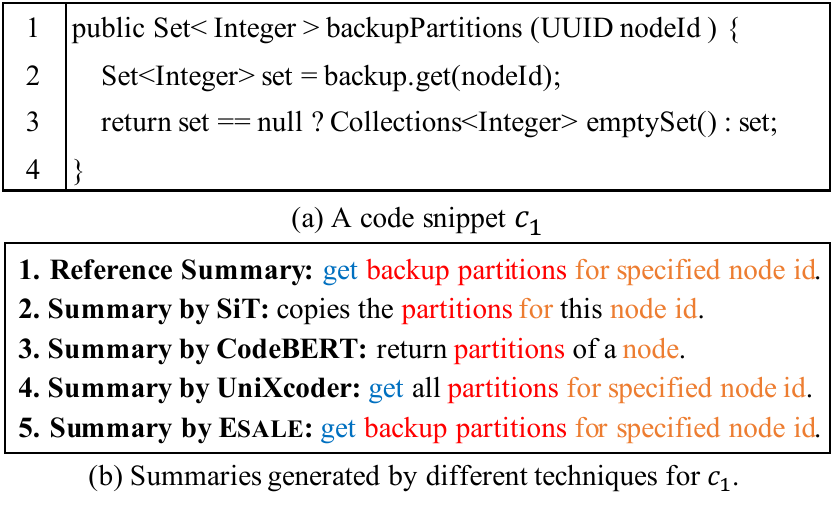}
  \caption{Code snippet $c_1$ and summaries generated by different techniques for $c_1$.\revise{This example is from the test set of the JCSD dataset, numbered 8,335.}}
  \label{fig:motivation_example_1}
\end{figure}

\deletion{In this section, we take}\revision{This section takes} a real-world code snippet and corresponding \deletion{summary}\revision{summaries} generated by different techniques as examples to illustrate our motivation. 
\deletion{We take the code snippet $c_1$ in \deletion{Figure}\revision{Fig.}~\ref{fig:motivation_example_1}(a) as an example and apply different techniques to generate summaries for comparison. It}\revision{The code snippet $c_1$ in Fig.~\ref{fig:motivation_example_1}(a)} is \deletion{an example }from the \revise{test set of the} JCSD dataset\revise{ (numbered 8,335)}~\cite{2018-TL-CodeSum} (see details in Section~\ref{subsubsec:dataset}). The first line of \deletion{Figure}\revision{Fig.}~\ref{fig:motivation_example_1}(b) shows the comment written by the developer for $c_1$. We consider the comment as a reference summary\deletion{ (i.e., ground-truth)}. According to the grammar rules in natural language, we can simply divide the reference summary into three parts: ``get'' (Blue font), ``backup partitions'' (Red font), and ``for specified node id'' (Orange font). Lines 2-5 show the summaries generated by baselines SiT~\cite{2021-SiT}, CodeBERT, UniXcoder, and our {\toolname}, respectively. SiT is one of the \deletion{state-of-the-art}\revision{advanced} code summarization techniques. CodeBERT\deletion{~\cite{2020-CodeBERT}} is a representative pre-trained model for source code. UniXcoder\deletion{~\cite{2022-UniXcoder}} is the state-of-the-art pre-trained model for source code. Both CodeBERT and UniXcoder can also be used for code summarization tasks by fine-tuning. More details on the three baselines are introduced in Section~\ref{subsubsec:result_of_RQ1}. From \deletion{Figure}\revision{Fig.}~\ref{fig:motivation_example_1}, it can be observed that\deletion{,} compared with the reference summary, 1) SiT and CodeBERT can cover some words in the second and third parts (i.e., ``partitions'' and ``node id''); 2) UniXcoder is better than SiT and CodeBERT, and can cover the word ``partitions'' in the second part and all words in the last part (i.e., ``for specified node id'')\deletion{.}\revision{;} 3) \deletion{our }{\toolname} performs the best, successfully covering all three parts. Compared to \deletion{the best baseline }UniXcoder, our {\toolname} can correctly generate the word ``backup''. 

\begin{figure}[htbp]
\centering
    \subfigure[UniXcoder]
    {
        \includegraphics[width=0.45\linewidth]{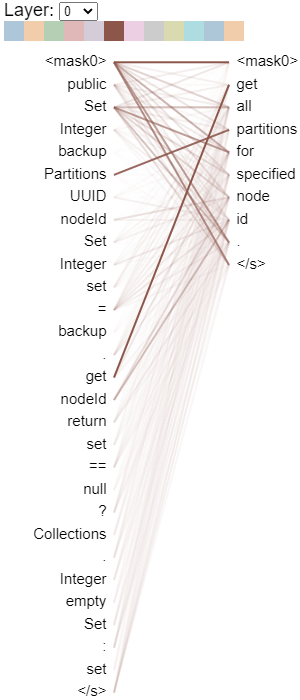}
        \label{fig:UniXcoder_layer0_head_5}
    }
    \subfigure[{\toolname}]
    {
        \includegraphics[width=0.45\linewidth]{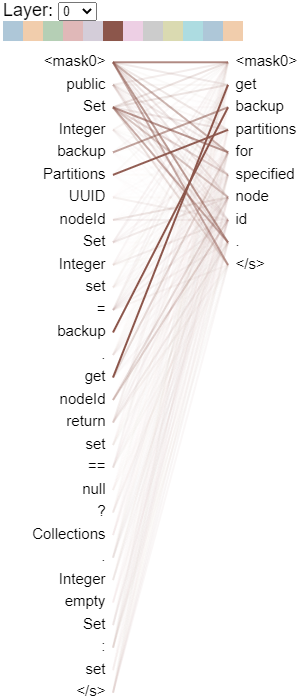}
        \label{fig:ESALE_layer0_head_5}
    }
    \caption{Visualization of cross attention}
    \label{fig:cross_attention_Visualization}
\end{figure}

To intuitively understand how \deletion{our technique}\revision{{\toolname}} can perform better in code summarization, we visualize the cross attention between encoder and decoder (also called encoder-decoder attention~\cite{2017-Transformer}) using the attention visualization tool BertViz\footnote{\url{https://github.com/jessevig/bertviz}}~\cite{2019-Analyzing-Attention-in-Transformer}. Attention can help interpret the model by showing how the model attends to different parts of the input~\cite{2015-NMT-Jointly-Learning-to-Align--Translate, 2019-Analysis-Methods-in-NLP-Survey, 2019-Analyzing-Attention-in-Transformer}. \deletion{Figure}\revision{Fig.}~\ref{fig:cross_attention_Visualization}(a) and (b) show the visualizations of the cross attentions of UniXcoder\deletion{,} and our {\toolname}. 
In \deletion{Figure}\revision{Fig.}~\ref{fig:cross_attention_Visualization}(a) and (b), the cross attention is depicted as lines connecting the attending tokens (right) with the tokens being attended to (left). In our setting, the left and right show code snippets and summaries, respectively. Colors identify the head(s), and the thickness of the line reflects the attention weights. \deletion{From \deletion{Figure}\revision{Fig.}~\ref{fig:cross_attention_Visualization}(a) and (b), it}\revision{It} can be observed that each word of the generated summary is basically mapped from a certain code pattern (a set of tokens with different weights). In addition, compared to UniXcoder, \deletion{our }{\toolname} can correctly predict the word ``backup'', which can be attributed to the higher attention {\toolname} paid to the two ``backup'' tokens in the code snippet. This example demonstrates that the context vector produced by {\toolname}'s encoder captures the code pattern that needs to be translated into ``backup''. It should be noted that the main difference between our {\toolname} and UniXcoder is the encoder. We initialize {\toolname}'s encoder with the parameters of the pre-trained encoder of UniXcoder, and then fine-tune it with three summary-focused tasks (details are described in Section~\ref{subsubsec:shared_encoder_training}). Hence, we can attribute the better performance of {\toolname} to its encoder. As well, we can boldly speculate that the three summary-focused tasks we designed could teach the encoder to learn the code-comment alignment and capture important code features that need to be translated. 

\begin{table}[htbp]
    \scriptsize
    \renewcommand{\arraystretch}{1.2}
    \tabcolsep=18pt
    \caption{\revision{Summaries generated by different models \revised{when the token ``backup'' is deleted from $c_1$}. {\toolname} w/o AWP, MLM, and ULM denote that we pre-train the shared encoder of {\toolname} without AWP, MLM, and ULM, respectively.}}
    \label{tab:motivation_case}
    \centering
    \begin{tabular}{|l|l|}
        \hline
        \revision{Model} & \revision{Summary} \\
        \hline
    
        \revision{UniXcoder} & \revision{get partitions for a given node id.} \\
        \hline
        
        \revision{{\toolname}} & \revision{get partitions for specified node id.} \\
        
        \hline
        \hline
        
        \revision{{\toolname} w/o AWP} & \revision{\# of partitions for given node id.} \\

        \hline
        
        \revision{{\toolname} w/o MLM} & \revision{\makecell[l]{provide this method to get the \\partitions for the given node id.}} \\

        \hline
    
        \revision{{\toolname} w/o ULM} & \revision{\# of partitions for given node id.} \\
        \hline
    \end{tabular}
\end{table}

\deletion{In order to}\revision{To} confirm our speculation, we further study the summaries generated by {\toolname} by deleting related code patterns in the motivation case. When deleting two related code tokens ``backup'' from \deletion{the code snippet }$c_1$ (Lines 1 and 2 of \deletion{Figure}\revision{Fig.}~\ref{fig:motivation_example_1}(a)), the summaries generated by UniXcoder and {\toolname} are\revision{ shown in the first two rows of TABLE~\ref{tab:motivation_case}.}\deletion{ as follows:
\begin{center}
\textbf{UniXcoder}: get partitions for a given node id.

\end{center}
}
It can be observed that compared with the respective original summaries shown in \deletion{the }lines 4 and 5 of \deletion{Figure}\revision{Fig.}~\ref{fig:motivation_example_1}(b), the new summaries generated by UniXcoder and {\toolname} have different changes. Specifically, the new summary generated by UniXcoder misses the word ``all''\revision{,} and ``a given node id" is different from ``specified node id'' in the reference summary.
For our {\toolname}, the word ``backup'' is not included in the new\revision{ly} generated summary because it is not present in the code any more. Even human developers cannot tell. It also proves in reverse that {\toolname}'s encoder can capture the code pattern related to the summary word ``backup'' when it appears in the input code snippet while UniXcoder's encoder cannot. All these can be attributed to the three summary-focused tasks we designed, on which the encoder trained is able to learn the code-comment alignment and capture important code features that need to be summarized. 
\revision{In addition, when the word ``backup'' is omitted, the summaries generated by {\toolname} trained without the AWP, MLM, or ULM task are shown in the last three rows of TABLE~\ref{tab:motivation_case}. It is observed that all three versions of {\toolname} do not generate the word ``backup''. It means that the three summary-focused tasks indeed help train {\toolname}'s encoder to capture the code pattern related to the summary word ``backup'' when it appears in the input code snippet while UniXcoder's encoder cannot. Certainly, it is undeniable that each task may influence the generation of other parts of the summary content, and the three tasks may also affect each other. It is important to note that this still serves to demonstrate that the superior performance of {\toolname} originates from the three summary-focused tasks.}

\section{Methodology}
\label{sec:methodology}

\subsection{Overview}
\label{subsec:overview}

\revision{Our approach produces a code summarization model via two sequential training phases: (a) training a shared encoder, followed by (b) training a code summarization model based on the encoder generated in phase (a). }

\deletion{\deletion{Figure}\revision{Fig.}~\ref{fig:overview_of_our_approach} illustrates the overview of our approach. Part (a) shows the training procedure of the shared encoder, and parts (b) and (c) show the training and usage of the code summarization model of {\toolname}, respectively. }
\revision{In phase (a), }{\toolname} decomposes the training procedure of the shared encoder into two \deletion{phases}\revision{steps}: preprocessing and shared encoder training. 
In the first \deletion{phase}\revision{step}, {\toolname} takes in pairs of code snippets and comments in the training data and produces two sequences, i.e., input sequences and masked input sequences\deletion{ (details are discussed in Section~\ref{subsubsec:preprocessing})}\revision{, detailed in Section~\ref{subsubsec:preprocessing}}. 
In the second \deletion{phase}\revision{step}, {\toolname} utilizes the two sequences to train a shared encoder. In this paper, we aim to enhance the encoder to learn the code-summary alignment, thereby improving code summarization performance. Therefore,\deletion{ in the second phase,} we exploit the \deletion{multi-task learning}\revision{MTL} paradigm to train a shared encoder on three summary-focused tasks. Specifically, for input sequences, {\toolname} utilizes a shared encoder to transform them into embedding representations\deletion{ (i.e., $\bm{e}^{I}$)}, which will be used in the AWP and ULM tasks. For masked input sequences, {\toolname} utilizes the shared encoder to transform them into embedding representations\deletion{ (i.e., $\bm{e}^{M}$)}, which will be used in the MLM task. The models for the three tasks are jointly trained and determine the parameters of the shared encoder\deletion{(details are discussed in Section~\ref{subsubsec:shared_encoder_training})}\revision{, detailed in Section~\ref{subsubsec:shared_encoder_training}}. 

\deletion{As shown in \deletion{Figure}\revision{Fig.}~\ref{fig:overview_of_our_approach}(b)}\revision{In phase (b)}, after getting the \deletion{well-trained}\revision{pre-trained} shared encoder, \deletion{we further fine-tune}\revision{{\toolname} fine-tunes} it and \deletion{train}\revision{trains} a decoder simultaneously on the \revision{downstream }code summarization task\deletion{ (details are discussed in Section~\ref{subsec:training_of_code_summarization_model})}. 
The fine-tuned shared encoder and fine-tuned decoder compose a well-trained code summarization model\revision{, detailed in Section~\ref{subsec:training_of_code_summarization_model}}. When the well-trained code summarization model is deployed online, it can take in a code snippet given by the developer and generate a natural language summary\deletion{. \deletion{Figure}\revision{Fig.}~\ref{fig:overview_of_our_approach}(c) shows the usage of the well-trained code summarization model, and details are discussed in Section~\ref{subsec:deployment_of_code_summarization_model}}\revision{, detailed in Section~\ref{subsec:deployment_of_code_summarization_model}}.

\subsection{Training of Shared Encoder}
\label{subsec:training_of_shared_encoder}
\deletion{As shown in the part (a) of \deletion{Figure}\revision{Fig.}~\ref{fig:overview_of_our_approach}, the training of the shared encoder is completed by two sequential phases. In this section, we detail the two phases.}

\subsubsection{Preprocessing}
\label{subsubsec:preprocessing}
\deletion{In the preprocessing \deletion{phase}\revision{step}, {\toolname} aims to process the raw data into the format required by the next \deletion{phase}\revision{step} (i.e., shared encoder training). Specifically, }{\toolname} takes in raw training data where each sample consists of a code snippet and its corresponding comment. {\toolname} follows common practices~\cite{2020-CodeBERT, 2021-GraphCodeBERT, 2022-UniXcoder} and uses the tokenizer provided by\revise{ Roberta}~\cite{2019-RoBERTa} to tokenize code snippets and comments and produce token sequences and word sequences, respectively\deletion{ (Step \circled{1} in \deletion{Figure}\revision{Fig.}~\ref{fig:overview_of_our_approach}(a))}. \revise{We also use \deletion{BPE}\revision{Byte Pair Encoding (BPE)} within Roberta\revision{ to split tokens into subtokens}.} As~\cite{2021-Why-My-Code-Summarization-Not-Work}, we call the basic unit of preprocessed source code a token and the basic unit of summary a word. {\toolname} further masks parts of words in word sequences to produce masked word sequences\deletion{ (Step \circled{2})}. Specifically, we follow existing works~\cite{2020-CodeBERT, 2022-UniXcoder} and randomly choose 15\% of the words in a word sequence, and change them into a special token $<$MASK$>$. Next, two special tokens $<$SOS$>$ and $<$EOS$>$ are added at the beginning and the end of \revision{the }token sequences, respectively. The special token $<$EOS$>$ is appended as a suffix for word sequences and masked word sequences. Then we concatenate pairs of token sequences and word sequences to produce input sequences\deletion{ (Step \circled{3})}. We concatenate pairs of token sequences and masked word sequences to produce masked input sequences\deletion{ (Step \circled{4})}. Unlike pre-trained models (e.g., CodeBERT and UniXcoder) that first concatenate pairs of token sequences and word sequences and then mask parts of the input sequences, we only mask parts of word sequences. Both input sequences and masked input sequences will be used to train the shared encoder in the second \deletion{phase}\revision{step}.

\subsubsection{Shared Encoder Training}
\label{subsubsec:shared_encoder_training}
\deletion{As mentioned earlier, we exploit the multi-task learning paradigm to train a shared encoder, including one domain-specific task (i.e., AWP) and two general tasks borrowed from the field of NLP (i.e., ULM and MLM). Although the three tasks use three different models, they share an encoder. Thus, we first introduce the design of the shared encoder.}

\textbf{Shared Encoder.}
The shared encoder is a deep neural network or pre-trained model responsible for transforming the input sequences into vector representations (i.e., embeddings). \deletion{For example, in the AWP task, the shared encoder transforms the input sequences into embeddings $\bm{e}^{I}$. }In practice, we build our shared encoder upon the existing pre-trained encoder. There are two benefits to doing this: 1) compared to training the encoder from scratch, the scheme based on the pre-trained encoders can significantly reduce the training cost; 2) existing pre-trained encoders have achieved almost optimal performance on the code summarization task, providing a high starting point. 

Specifically, we tried to build our shared encoder upon two pre-trained encoders provided by CodeBERT and UniXcoder. We first initialize the shared encoder with the parameters of their pre-trained encoders. Then, we fine-tune the shared encoder with three summary-focused tasks, i.e., AWP, ULM, and MLM. The experimental results show that the shared encoder built upon UniXcoder's encoder is better than that built on CodeBERT's encoder on the downstream code summarization task\deletion{ (details are discussed in Section~\ref{subsubsec:result_of_RQ1})}\revision{, detailed in Section~\ref{subsubsec:result_of_RQ1}}.

Next, we introduce the design of the \deletion{tree}\revision{three} summary-focused tasks.

\textbf{(i) \deletion{Action Word Prediction}\revision{AWP}.} 
\revision{An ``action word'' in a summary is typically a verb that broadly classifies what the code does~\cite{2021-Action-Word-Prediction-for-Code-Summarization}, such as ``get'', ``add'', and ``remove''. Programmers tend to write summaries containing only one action word, typically positioned at the beginning of the summary sentence (i.e., the first word).}

\deletion{The AWP task is defined as the problem of predicting the action word to be used in the summary given a code snippet~\cite{2021-Action-Word-Prediction-for-Code-Summarization}. }
\revision{AWP is a classification task, where the input of the model is a code snippet, and the output is the predicted label with respect to the action word~\cite{2021-Action-Word-Prediction-for-Code-Summarization}.}
\deletion{Therefore, in this paper, this}\revision{In this paper, we use this} task \deletion{is designed }to train a model capable of predicting the action words of summaries based on given code snippets. 
\revision{Formally, let $c = \{t_1, t_2, \dots, t_m\}$ denote the token sequence of the code snippet, where $m$ is the length of the token sequence, and $y = \{y_1, y_2, \dots, t_C\}$ denote the set of possible classes, where $C$ is the number of classes of action words. The summary-focused AWP can be defined as follows:
\begin{definition}[Summary-focused AWP]
    A summary-focused AWP is a multi-classification task denoted as $\hat{y} = \arg\max_{y \in Y} P(y|c)$, where:
    \begin{itemize}
        \item $P(y|c)$ represents the probability of the class $y$ given the code snippet $c$. 
        \item $\arg\max$ denotes the operation that selects the class label with the highest probability.
    \end{itemize}
\end{definition}
}

\deletion{Such a model}\revision{The model we train} is composed of the shared encoder and a classification layer. The classification layer is a \deletion{full connect}\revision{fully connected} network of size $N * \delete{41}\revise{C}$, where $N$ is the output size of the shared encoder\deletion{\revise{ and $C$ is the number of classes of action words}}. Given an input sequence $x$, we first utilize the shared encoder to transform $x$ into the embedding $\bm{e}^x$\deletion{ (Step \circled{5})}. Then, a classification layer is used to classify $\bm{e}^x$ into predicted action word classes\deletion{ (Step \circled{7})}. Given the embedding $\bm{e}^x$, we obtain the logits by $\hat{y}_i = \bm{W}\bm{e}^x+\bm{b}$, where $\bm{W}$ is the weight matrix and $\bm{b}$ is the bias term. We optimize the model by minimizing the categorical cross-entropy loss:
\begin{equation}
    \mathcal{L}_{AWP}(\Theta) = -\sum_{i=1}^{C}y_i{\log\frac{\exp(\hat{y}_i)}{\sum_{j=1}^{C}{\exp(\hat{y}_{j})}}}
    \label{equ:loss_AWP}
\end{equation}
where $\Theta$ denotes trainable parameters of the model (i.e., $\bm{W}$ and $\bm{b}$); $\hat{y}_i$ and $y_i$ are the predicted score and target score for each class $i \in C$\deletion{; $C$ is the number of classes of action words}. In practice, we follow\deletion{ the work}~\cite{2021-Action-Word-Prediction-for-Code-Summarization} and use the top-40 setting; that is, the model attempts to predict the forty most-common action words, or ``other'' if predicted to be a less-common action word. The top 40 action words are selected based on their frequency in the comments of all samples in each dataset.

\revise{Here, we give a brief explanation of why we consider AWP as one of the pre-training tasks. \deletion{As mentioned in Section~\ref{subsec:action_word_prediction},}\revision{First,} the production of good summaries relies on the production of the action words in those summaries. \deletion{However, current code summarization techniques already try to predict the action word along with the whole summary, and yet action word prediction on its own is quite difficult.}\revision{ Code summarization models are widely built on the encoder-decoder framework, where the decoder predicts words one by one according to previous words and the context produced by the encoder. So, if the first word is wrong, it is difficult for the decoder to predict the entire summary correctly. This situation can be exaggerated by the aggressive use of attention mechanisms, which can attend previous words in the predicted summary to parts of the code snippet~\cite{2021-Action-Word-Prediction-for-Code-Summarization}. Therefore, it is crucial for code summarization models to predict accurate action words. Second, our experiments found that {\toolname} equipped with AWP performs better than without.} 
In practice, before deciding to add AWP as one of the pre-training tasks, we also followed~\cite{2021-Action-Word-Prediction-for-Code-Summarization} and conducted experiments on the performance of encoders of different seq2seq models on the AWP task with 41 classes. \deletion{Specifically, }We compared five techniques, including AttendGRU, CodeBERT, UniXcoder, {\toolname} w/o AWP, and {\toolname}. AttendGRU is representative of seq2seq-like approaches as proposed by Iyer et al.~\cite{2016-CODE-NN}. In the paper~\cite{2021-Action-Word-Prediction-for-Code-Summarization}, AttendGRU performs the best, so we also consider it as a baseline. For AttendGRU, we build a classification model by appending a \deletion{full connect}\revision{fully connected} network to its encoder, and train the model from scratch.\deletion{ CodeBERT and UniXCoder are two pre-trained models of code (detailed in Section~\ref{subsubsec:result_of_RQ1}). {\toolname} w/o AWP means that we train the shared encoder of {\toolname} without the AWP task.} For CodeBERT, UniXcoder, {\toolname} w/o AWP, and {\toolname}, we build classification models by appending a \deletion{full connect}\revision{fully connected} network to their \deletion{pre-trained }encoders as classification layers and train the models by fine-tuning.

TABLE~\ref{tab:compare_action_word_prediction} shows the experimental results where columns 2--4 report the weighted average precision, recall, and f-measure computed by the classification\_report function provided by scikit-learn~\footnote{\url{https://scikit-learn.org/stable/modules/model_evaluation.html\#classification-report}}. The experimental dataset consists of pairs of code snippets and action words extracted from the JCSD dataset\deletion{ (details of the dataset introduced in Section~\ref{subsubsec:dataset})}. From the table, it is observed that, overall, 1) compared with the encoder from AttendGRU and trained from scratch, the pre-trained encoders from CodeBERT, UniXcoder, {\toolname} w/o AWP perform better in terms of \deletion{F}\revision{f}-measure; 2) compared with {\toolname} w/o AWP, {\toolname}'s encoder treating AWP as one of the pre-training tasks further improves the \revision{AWP }performance\deletion{ of action word prediction}. More details of comparing {\toolname} w/o AWP and {\toolname} are described in Section~\ref{subsubsec:results_of_RQ2}.

In summary, \deletion{we use the AWP task as a pre-training task, aiming}\revision{equipping {\toolname} with AWP aims} to enhance the shared encoder to learn the code pattern that is the key feature to predict the action word. In this way, the code summarization model based on the shared encoder can better generate the action word of the summary.
}

\begin{table}[htbp]
  \small
  \renewcommand{\arraystretch}{1.2}
  \caption{\revise{The performance of encoders of different seq2seq models on the AWP task. {\toolname} w/o AWP denotes that we pre-train the shared encoder of {\toolname} without AWP.}}
  \label{tab:compare_action_word_prediction}
  \centering
  \begin{tabular}{|l|c|c|c|}
    \hline
   \revise{Model (Year)} & \revise{Precision} & \revise{Recall} & \revise{F-measure} \\
    \hline
    \revise{AttendGRU (2016)} & \revise{63.49} & \revise{62.43} & \revise{62.55} \\
    \hline
    \revise{CodeBERT (2020)} & \revise{63.17} & \revise{66.01} & \revise{62.65}\\
    \revise{UniXcoder (2022)} & \revise{63.19} & \revise{66.28} & \revise{62.87}\\
    \hline
    \hline
    \revise{{\toolname} w/o AWP} & \revise{63.17} & \revise{66.21} & \revise{63.05}\\
    \revise{{\toolname}} & \revise{63.54} & \revise{66.26} & \revise{63.27}\\
    \hline
 \end{tabular}
\end{table}

\textbf{(ii) \deletion{Unidirectional Language Modeling}\revision{ULM}.} The ULM task is defined as the problem of left-to-right language modeling~\cite{2018-Unidirectional-Language-Modeling}, which only allows words to attend the previous words and itself to predict the next word~\cite{2022-UniXcoder, 2020-Multi-task-Learning-Code-Completion}. 
\revision{Unlike existing works on predicting the next code token~\cite{2022-UniXcoder, 2020-Multi-task-Learning-Code-Completion}, we use this task to train a model capable of predicting the next summary word one by one conditioned on the code token sequence and unmasked\revision{/preceding} parts of the summary sequence. It can be done using a ULM mask matrix for the attention mask.}
\revision{We refer to such ULM as summary-focused ULM. Formally, $w = \{w_1, w_2, \dots, w_n\}$ denote the word sequence of the summary, where $n$ is the length of the word sequence. The summary-focused ULM can be defined as follows:
\begin{definition}[Summary-focused ULM]
    A summary-focused ULM is a probabilistic model denoted as $P(w_1, w_2, \dots, w_n|c) = \prod_{i=1}^{n}P(w_i|w_{i-1}, w_{i-2}, \dots, w_1, c)$, where:
    \begin{itemize}
        \item $P(w_i|w_{i-1}, w_{i-2}, \dots, w_1, c)$ represents the probability of the word $w_i$ given the preceding summary words $w_{i-1}, w_{i-2}, \dots, w_1$ and the code snippet $c$. 
        \item $\prod$ denotes the product of probabilities over the entire word sequence of the summary.
    \end{itemize}
\end{definition}
}
\deletion{Therefore, in this paper, this task is designed to train a model capable of predicting the next word based on the previous words. }\deletion{Such a model}\revision{The model we train} also includes the shared encoder followed by a \deletion{full connect}\revision{fully connected} network. The size of the \deletion{full connect}\revision{fully connected} network is $N * |V|$, where $N$ is the output size of the shared encoder and $|V|$ is the vocabulary size. Given an input sequence $x$, we first utilize the shared encoder to produce its embedding $\bm{e}^x$\deletion{ (Step \circled{5})}. Then, the \deletion{full connect}\revision{fully connected} network is used to predict the likelihood score of each word in the vocabulary being the next word\deletion{ (Step \circled{8})}. The model is optimized by minimizing the objective function:
\begin{equation}
	\mathcal{L}_{ULM}(\Theta) = - \sum_{i=0}^{n-1}logp(w_i|\bm{e}^{x}_{t<i}).
	\label{equ:loss_ULM}
\end{equation}
where $\bm{e}^{x}_{t<i}$ represents the embedding of the word sequence appearing on the left of the word $w_i$.

\begin{figure}[htbp]
  \centering
  \includegraphics[width=\linewidth]{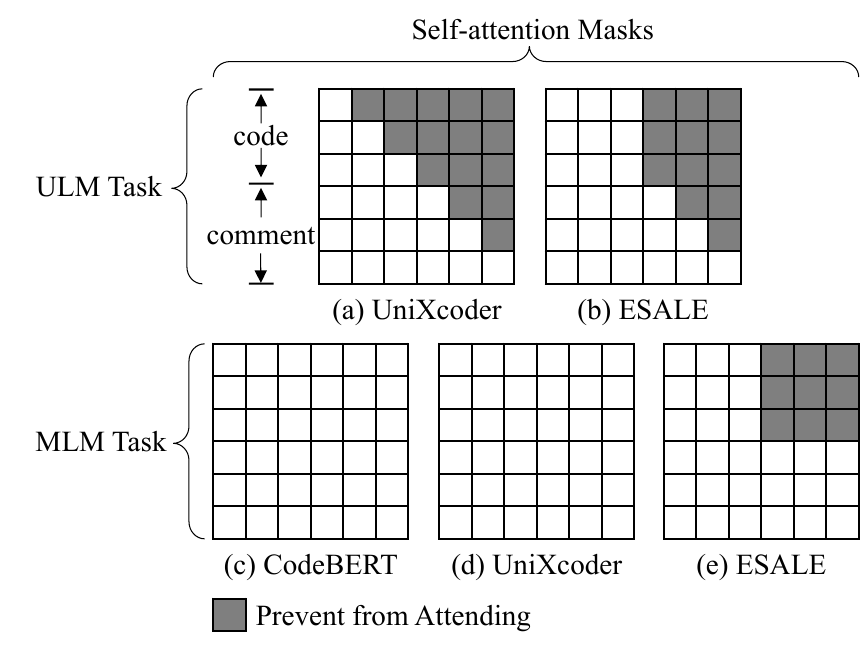}
  \caption{\revise{Differences in preventing from attending}}
  \label{fig:difference_in_mask}
\end{figure}

\deletion{Unlike existing works on predicting the next code token~\cite{2022-UniXcoder, 2020-Multi-task-Learning-Code-Completion}, we use this task to train a model capable of predicting the next summary word one by one conditioned on the code token sequence and unmasked parts of the summary sequence. It can be done using a ULM mask matrix for the attention mask, as shown in \deletion{Figure}\revision{Fig.}~\ref{fig:overview_of_our_approach}(a).} \revise{Fig.~\ref{fig:difference_in_mask}(a) and (b) visually illustrate the self-attention masks used by UniXcoder and {\toolname}, respectively. The self-attention masks are used to control the behavior of the model, i.e., preventing from attending. UniXcoder directly exploits the general ULM in NLP~\cite{2019-Unified-Language-Model-for-NLP}, which uses a triangular matrix for attention mask, predicting the next token in the entire input sequence. Different UniXcoder, {\toolname} introduces a summary-focused ULM, which is used to train {\toolname} to predict the next summary word only in the summary word sequence by attending the entire code token sequence and the left summary words.}

\textbf{(iii) \deletion{Masked Language Modeling}\revision{MLM}.}
The MLM task is defined as the problem of predicting the original tokens of masked tokens based on their bidirectional contextual tokens~\cite{2019-BERT}. \revision{Unlike ULM, which can only be trained unidirectionally, bidirectional conditioning in MLM allows each word to indirectly see itself, simplifying the prediction of the target word in a multi-layered context.} 
Therefore, in this paper, this task is designed to train a model capable of predicting masked tokens based on all tokens in the code snippet and unmasked words in the summary. 
\revision{Similarly, we refer to such MLM as summary-focused MLM. Formally, the summary-focused MLM can be defined as follows:
\begin{definition}[Summary-focused MLM]
    A summary-focused MLM is a probabilistic model denoted as $P(w_1, w_2, \dots, w_n|c) = \prod_{i=1}^{n}P(w_i|w_1, \dots, w_{i-1}, w_{i+1}, \dots, w_{n}, c)$, where:
    \begin{itemize}
        \item $P(w_i|w_1, \dots, w_{i-1}, w_{i+1}, \dots, w_{n}, c)$ represents the probability of the masked word $w_i$ given the unmasked summary words $w_1, \dots, w_{i-1}, w_{i+1}, \dots, w_{n}$ and the code snippet $c$. 
        \item $\prod$ denotes the product of probabilities over the entire word sequence of the summary.
    \end{itemize}
\end{definition}
}

\deletion{Such a model}\revision{In this task, the model we train} is composed of a shared encoder and a \deletion{full connect}\revision{fully connected} network. 
\deletion{The size of the \deletion{full connect}\revision{fully connected} network is $N * |V|$. $N$ is the output size of the shared encoder and $|V|$ is the Vocabulary size. }\revision{The design of the fully connected network is the same as in the summary-focused ULM task.}
Given a masked input sequence $x$, we first use the shared encoder to transform $x$ into embedding $\bm{e}^x$. Then, the \deletion{full connect}\revision{fully connected} network is used to predict the likelihood score of each word in the vocabulary being the masked word\deletion{ (Step~\circled{9})}. We optimize the model by minimizing the following objective function:
\begin{equation}
	\mathcal{L}_{MLM}(\Theta) = -\sum_{w_i \in S_m}{logp(w_i|\bm{e}^x_{mask})}
	\label{equ:loss_MLM}
\end{equation}
where $\bm{e}^x_{mask}$ is the embedding of \deletion{the masked input sequence}\revision{$x$}; $S_m$ is the set of masked words that need to be predicted.

\begin{figure*}[htbp]
  \centering
  \includegraphics[width=\linewidth]{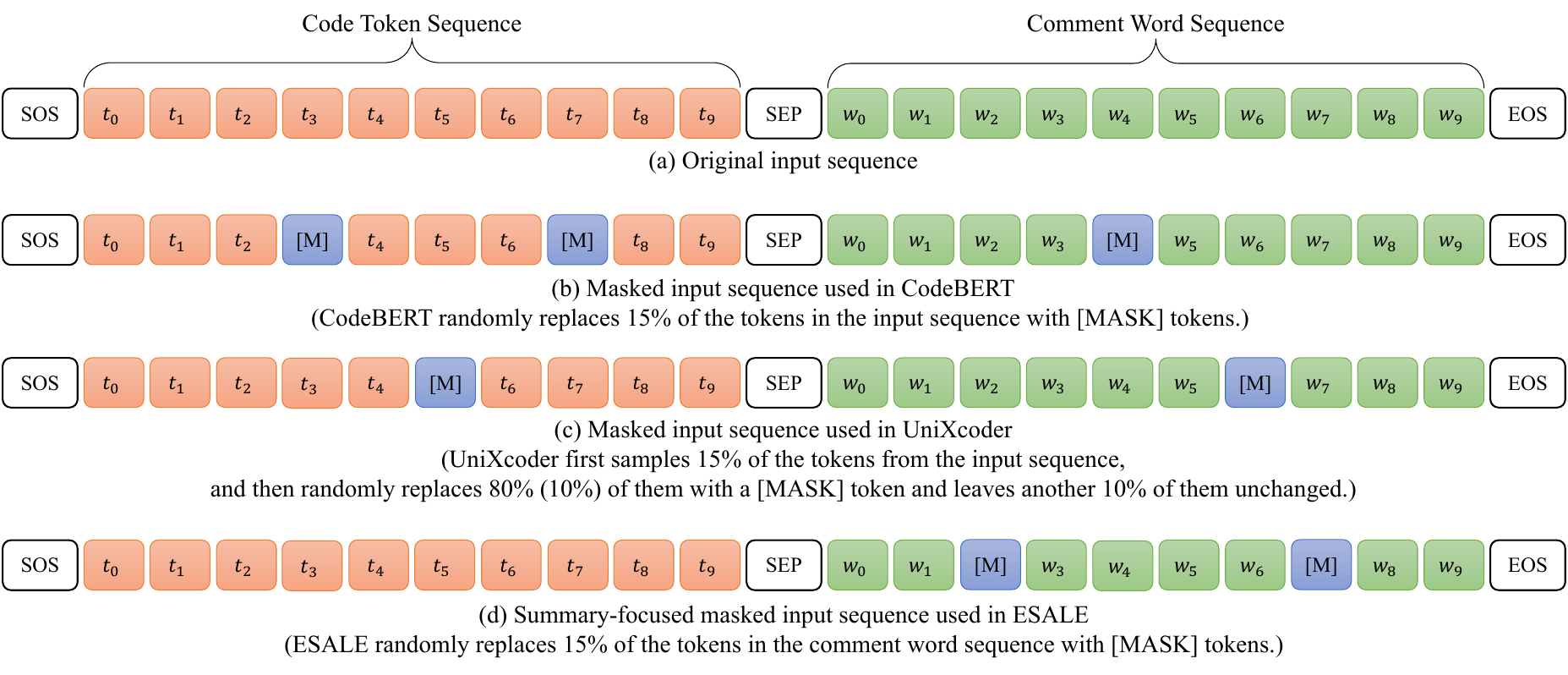}
  \caption{\revise{Differences in MLM between baselines (CodeBERT, UniXcoder) and our {\toolname}}}
  \label{fig:difference_in_MLM}
\end{figure*}

\revise{We use Fig.~\ref{fig:difference_in_MLM} to visually illustrate the differences in the masked proportion and \deletion{positions}\revision{position} between \revision{the }baselines (i.e., CodeBERT and UniXcoder) and our {\toolname}. Fig.~\ref{fig:difference_in_MLM}(a) shows an example of an original input sequence consisting of a code token sequence and a comment word sequence. Fig.~\ref{fig:difference_in_MLM}(b)--(d) show the masked input sequences used in CodeBERT, UniXcoder, and {\toolname}, respectively. In Fig.~\ref{fig:difference_in_MLM}(b), we follow CodeBERT and randomly replace 15\% of the tokens in the input sequence with [MASK] tokens (the blue blocks labeled [M] in the figure). In Fig.~\ref{fig:difference_in_MLM}(c), we follow UniXcoder and first sample 15\% of the tokens from the input sequence, and then randomly replace 80\% (i.e., about 10\% of the input sequence) of them with a [MASK] token and leave another 10\% of them unchanged. In Fig.~\ref{fig:difference_in_MLM}(d), our {\toolname} randomly replaces 15\% of the tokens in the comment word sequence with [MASK] tokens. From the figure, it is observed that {\toolname} is summary-focused and significantly different from CodeBERT and UniXcoder in the masked proportion and \deletion{positions}\revision{position}. Fig.~\ref{fig:difference_in_mask}(c), (d), and (e) also visually present the self-attention masks used by CodeBERT, UniXcoder, and {\toolname}, respectively.}

\revision{\textbf{Model Training. }}The training procedure of the above \revision{task} models follows the existing \deletion{multi-task learning (MTL)}\revision{MTL} paradigm~\cite{2020-Multi-task-Learning-Code-Completion}. \revision{In MTL, models are trained with data from multiple related tasks simultaneously while using a shared representation to learn the common features between all these tasks, and what is learned for one task can help other tasks be learned better~\cite{1997-Multitask-Learning}. The shared representation increases data efficiency and can potentially yield a faster learning speed for related or downstream tasks, helping to alleviate the well-known weaknesses of deep learning: large-scale data requirements and computational demand~\cite{2020-Multi-Task-Learning-Survey}. 
}

\revision{In this paper, we exploit the MTL paradigm to train a shared encoder with three summary-focused tasks, i.e., AWP, ULM, and MLM.}
The weight parameters of the shared encoder are learned to minimize the sum of the cross-entropy losses of the three pre-training tasks, and are shared among all three tasks. The final loss function is computed as:
\begin{equation}
	\mathop{\min}_{\Theta} \mathcal{L}_{AWP}(\Theta) + \mathcal{L}_{ULM}(\Theta) + \mathcal{L}_{MLM}(\Theta)
	\label{equ:loss_all}
\end{equation}

\revise{ 
Intuitively, during \deletion{pretraining}\revision{pre-training}, the AWP model predicts the label corresponding to action words based on the input sequence. Simultaneously, the ULM model predicts the next token based on the left tokens of the input sequence. Meanwhile, the MLM model predicts the original tokens of masked tokens of the input sequence. The AWP model, ULM model, and MLM model share an encoder (i.e., the encoder of ESALE). After obtaining a pre-trained encoder (referred to as the \deletion{well-trained}\revision{pre-trained} shared encoder in this paper), we further use pairs of code snippets and \deletion{comments}\revision{summaries} to fine-tune it along with the decoder. The fine-tuning process is elaborated upon in the subsequent section.}

\subsection{Training of Code Summarization Model}
\label{subsec:training_of_code_summarization_model}
After obtaining the \deletion{well-trained}\revision{pre-trained} shared encoder, we further train a code summarization model capable of generating a succinct natural language summary for a given code snippet. Specifically, we fine-tune the \deletion{well-trained}\revision{pre-trained} shared encoder on the code summarization task and simultaneously train a decoder. \deletion{\deletion{Figure}\revision{Fig.}~\ref{fig:overview_of_our_approach}(b) shows the training procedure of the code summarization model. }Given training data consisting of pairs of code snippets and comments, {\toolname} first leverages the \deletion{well-trained}\revision{pre-trained} shared encoder to transform code snippets into context vectors $\bm{e}^{Code}$. Then, {\toolname} leverages a decoder to generate predicted summaries. The decoder takes in $\bm{e}^{Code}$ and predicts words one by one\deletion{ (details are discussed}\revision{, detailed} in Section~\ref{subsubsec:decoder}). Finally, {\toolname} computes the loss ($\mathcal{L}_{CS}(\Theta)$) based on predicted summaries and ground-truth summaries (i.e., comments) and iteratively updates the model parameters $\Theta$\deletion{ (details are discussed}\revision{, detailed} in Section~\ref{subsubsec:model_training}).

\subsubsection{Decoder}
\label{subsubsec:decoder}
In this step, we utilize the decoder to generate natural language summaries. The decoder takes in the context vectors $\bm{e}^{Code}$ and predicts words one by one. Specifically, the decoder based on a neural network (e.g., LSTM~\cite{1997-LSTM} and Transformer~\cite{2017-Transformer}) is to unfold the context vectors $\bm{e}^{Code}$ into the target sequence (i.e., the word sequence of the summary) through the following dynamic model,
\begin{equation}
  \begin{aligned}
      \bm{h}_t = f(y_{t-1}, \bm{h}_{t-1}, \bm{e}^{Code})\\ p(y_t|Y_{<t}, X) = g(y_{t-1}, \bm{h}_t, \bm{e}^{Code})
  \end{aligned}
  \label{equ:sequence_prediction}
\end{equation}
where $f(\cdot)$ and $g(\cdot)$ are activation functions, $\bm{h}_t$ is the hidden state of the neural network at time $t$, $y_t$ is the predicted target word at $t$ (through $g(\cdot)$ with $Y_{<t}$ denoting the history $\{y_1, y_2, \cdots , y_{t-1}\}$. The prediction process is typically a classifier over the vocabulary. It can be seen from Equation~\revision{(}\ref{equ:sequence_prediction}\revision{)} that the probability of generating the target word is related to the current hidden state, the history of the target sequence\revision{,} and the context vectors $\bm{e}^{Code}$. The essence of the decoder is to classify the vocabularies by optimizing the loss function in order to generate the vector representing the feature of the target word $y_t$. After the vector passes through a \textit{softmax} function, the word corresponding to the highest probability is the result to be output.

In practice, we design our decoders with different schemes suggested by CodeBERT and UniXcoder, respectively. CodeBERT only provides a pre-trained encoder, while UniXcoder provides a pre-trained encoder and a pre-trained decoder. Therefore, when the shared encoder is built upon the pre-trained encoder provided by CodeBERT, we build our decoder upon Transformer~\cite{2017-Transformer}.  
When the shared encoder is built upon the pre-trained encoder provided by UniXcoder, we build our decoder upon the pre-trained decoder provided by UniXcoder.

\subsubsection{Model Training}
\label{subsubsec:model_training}
During the training of the code summarization model, the two components (\deletion{well-trained}\revision{pre-trained} shared encoder and decoder) are jointly trained to minimize the following objective function:
\begin{equation}
	\mathcal{L}_{CS}(\Theta) = -\frac{1}{N}\sum_{n=1}^{N}logp(\bm{y}_n|\bm{x}_n)
	\label{equ:loss_ab}
\end{equation}
where $\Theta$ is the model parameters of the code summarization model, and each $(\bm{x}_n,\bm{y}_n)$ is a (code snippet, comment) pair from the training set.

\subsection{Deployment of Code Summarization Model}
\label{subsec:deployment_of_code_summarization_model}
After the model is trained, we can deploy it online for code summarization service.\deletion{ Part (c) of \deletion{Figure}\revision{Fig.}~\ref{fig:overview_of_our_approach} shows the deployment of the code summarization model. }
For a code snippet $c$ given by the developer, {\toolname} first uses the fine-tuned encoder to transform $c$ into a context vector, which will be fed to the fine-tuned decoder to generate a summary in natural language. In practice, we can consider the well-trained {\toolname} as a black-box tool that takes in a code snippet \deletion{given by the developer }and generates a succinct natural language summary.

\section{Evaluation}
\label{sec:evaluation}
To evaluate our approach, in this section, we aim to answer the following four research questions:
\begin{description}
    \item[\textbf{RQ1:}] How does {\toolname} perform compared to the state-of-the-art baselines?
    \item[\textbf{RQ2:}] How do the three pre-trained tasks (i.e., AWP, ULM, and MLM) affect the performance of {\toolname} (ablation study)?
    \item[\textbf{RQ3:}] How does the robustness of {\toolname} perform when varying the code length and comment length?
    \item[\textbf{RQ4:}] How does {\toolname} perform in human evaluation?
\end{description}

\subsection{Experimental Setup}
\label{subsubsec:experimental_setup}

\subsubsection{Dataset}
\label{subsubsec:dataset}
\begin{table}[htbp]
  \scriptsize
  \tabcolsep=3pt
  \renewcommand{\arraystretch}{1.2}
  \caption{Dataset statistics. \revise{CPJD denotes the cross-project Java dataset. CSN denotes the CodeSearchNet corpus.}}
  \label{tab:statistics_of_datasets}
  \centering
  \begin{tabular}{|l|c|c|c|c|c|}
    \hline
    \multicolumn{2}{|c|}{Dataset} & Training Set & Validation Set & Test Set & \revision{Splitting Method} \\
    \hline
    \multicolumn{2}{|c|}{JCSD} & 69,708 & 8,714 & 8,714 & \revision{Random} \\
    \multicolumn{2}{|c|}{PCSD} & 57,203 & 19,067 & 19,066 & \revision{Random} \\
    \hline
    \multicolumn{2}{|c|}{\revise{CPJD}} & \revise{51,408} & \revise{7,180} & \revise{7,409} & \revision{Project-partitioned} \\
    \hline
    \multirow{4}{*}{CSN} & 
    \revise{PHP} & \revise{241,241} & \revise{12,982} & \revise{14,014} & \multirow{4}{*}{\revision{Project-partitioned}}\\
    & \revise{Go} & \revise{167,288} & \revise{7,325} & \revise{8,122} & \\
    & \revise{JavaScript} & \revise{58,025} & \revise{3,885} & \revise{3,291} & \\
    & \revise{Ruby} & \revise{24,927} & \revise{1,400} & \revise{1,261} & \\
    \hline
 \end{tabular}
\end{table}

\delete{We conduct experiments on two widely used datasets, including a Java dataset~\cite{2018-TL-CodeSum} (named JCSD~\cite{2020-Rencos}) and a Python dataset~\cite{2017-Corpus-Python-Functions} (named PCSD~\cite{2020-Rencos}), which}\revise{We conduct experiments on four datasets. JCSD provided by Hu et al.~\cite{2018-TL-CodeSum} is a Java dataset. PCSD provided by Barone et al.~\cite{2017-Corpus-Python-Functions} is a Python dataset. These two datasets are named JCSD and PCSD by Zhang et al.~\cite{2020-Rencos} and }have been widely used by existing code summarization studies~\cite{2019-Code-Generation-Summarization, 2020-CodeBERT, 2020-Transformer-based-Approach-for-Code-Summarization, 2020-Rencos, 2021-SiT, 2022-SCRIPT}. 
\deletion{We directly use the JCSD and PCSD datasets released by\deletion{ the state-of-the-art baseline}~\cite{2021-SiT}, which has split training/validation/test sets. }\revise{Some studies \deletion{pointed}\revision{point} out that randomly \deletion{split}\revision{splitting} datasets can lead to leakage between the test set and training set, since code from the same projects tends to be very similar~\cite{2019-Datasets-for-Code-Summarization, 2019-Adverse-Effects-of-Code-Duplication, 2022-Impact-Evaluation-on-Code-Summarization}. Therefore, we also conduct experiments on a recently released project-partitioned Java dataset (CPJD, for short) provided by Nie et al.~\cite{2022-Impact-Evaluation-on-Code-Summarization}.
To explore the performance of {\toolname} on multiple programming language datasets, we also conduct experiments on the CodeSearchNet corpus (CSN, for short). The CSN corpus provided by Husain et al.~\cite{2019-CodeSearchNet-Challenge} contains a large number of pairs of code snippets and comments across six programming languages, including Go, Java, JavaScript, PHP, Python, and Ruby. Lu et al.~\cite{2021-CodeXGLUE} \deletion{showed}\revision{reveal} that some comments contain content unrelated to the code snippets and performed data cleaning on the CSN corpus. Therefore, in this paper, we follow~\cite{2020-CodeBERT, 2021-CodeT5} and use the clean version of the CSN corpus provided by Lu et al.~\cite{2021-CodeXGLUE}. Since JCSD, PCSD, and CPJD already contain Java and Python datasets, \revised{to reduce the experimental workload and focus on evaluating {\toolname} on different languages, }for CSN, we mainly conduct experiments on four languages: PHP, Go, JavaScript, and Ruby.}
The statistics of the \delete{two}\revise{four} datasets are shown in \deletion{Table}\revision{TABLE}~\ref{tab:statistics_of_datasets}\revision{, where the last column presents the data splitting methods used by the four datasets.} 
\revise{It should be noted that we only use the training set in the pre-training phase of the shared encoder.}
\revision{Model training/pre-training on different datasets/programming languages is independent of each other.}

\subsubsection{Evaluation Metrics}
\label{subsubsec:evaluation_metrics}
We use three automatic metrics BLEU~\cite{2002-BLEU}, METEOR~\cite{2005-METEOR}, and ROUGE~\cite{2004-ROUGE}, to evaluate the model, which are widely used in code summarization~\cite{2022-SCRIPT, 2021-SiT, 2018-Improving-Code-Summarization-via-DRL, 2018-TL-CodeSum, 2017-Transformer, 2016-CODE-NN}. 

\textbf{BLEU}, the abbreviation for BiLingual Evaluation Understudy~\cite{2002-BLEU}, is widely used for evaluating the quality of generated summaries~\cite{2018-Improving-Code-Summarization-via-DRL, 2018-TL-CodeSum, 2016-CODE-NN}. It is a variant of precision metric, which calculates the similarity by computing the n-gram precision of a generated summary to the reference summary. It has a penalty for the overly short length~\cite{2002-BLEU}. 
In this paper, we follow~\cite{2021-SiT, 2022-SCRIPT, 2022-UniXcoder} and show \delete{the BLEU-4 score, which is often of interest as it can reflect the weighted results from 1 through 4}\revise{the standard BLEU score which provides a cumulative score of 1-, 2-, 3-, and 4-grams}~\cite{2021-Why-My-Code-Summarization-Not-Work}. 

\textbf{METEOR}, the abbreviation for Metric for Evaluation of Translation with Explicit ORdering~\cite{2005-METEOR}, is also widely used to evaluate the quality of generated summaries~\cite{2021-Code-Summarization-for-Smart-Contracts, 2020-Rencos, 2020-RL-Guided-Code-Summarization}. For a pair of summaries, METEOR creates a word alignment between them and calculates the similarity scores.

\textbf{ROUGE-L.} ROUGE is the abbreviation for Recall-oriented Understudy for Gisting Evaluation~\cite{2004-ROUGE}. ROUGE-L, a variant of ROUGE, is computed based on the longest common subsequence
(LCS). ROUGE-L is also widely used to evaluate the quality of generated code summaries~\cite{2021-Project-Level-Encoding-Code-Summarization, 2021-BASTS, 2021-API2Com}.

The scores of BLEU, METEOR, and ROUGE-L are in the range of [0, 1] and \revision{are }usually reported in percentages. The higher the scores, the closer the generated summary is to the reference summary, and the better the code summarization performance.

\subsubsection{Experimental Settings}
\label{subsubsec:experimental_settings}
To train models, we first shuffle the training data and set the mini-batch size to 32. For each batch, the code snippets are padded with a special token $\langle PAD \rangle$ to the maximum length. 
Following~\cite{2020-CodeBERT, 2022-UniXcoder}, we set the maximum length of code snippets and comments to 256 and 128, respectively.
We update the parameters via AdamW optimizer~\cite{2015-Adam} for 100k steps, with a learning rate of 0.0005. To prevent over-fitting, we use dropout with 0.1. For beam search, we set the beam size to 5. Finally, we select the best model based on the lowest validation loss. 
All models are implemented using the PyTorch 1.7.1 framework with Python 3.8. All experiments are conducted on a server equipped with one Nvidia Tesla V100 GPU with 32 GB memory, running on Ubuntu 18.04.

\subsection{Experimental Results}
\label{subsec:experimental_results}

\subsubsection{\textbf{RQ1:} {\toolname} vs. Baselines}
\label{subsubsec:result_of_RQ1}
\indent1)\textit{\;Baselines:}

\textbf{Re$^2$Com~\cite{2020-R2Com}} adopts an LSTM-based encoder-decoder architecture with \revision{an }attention mechanism. It first uses an information retrieval technique to retrieve a similar code snippet and treat its comment as an exemplar. Then, it uses \deletion{a}\revision{an} LSTM-based seq2seq neural network that takes the given code, its AST, its similar code, and its exemplar as input, and leverages the information from the exemplar to generate summaries.

\textbf{SiT~\cite{2021-SiT}} adopts a Transformer-based encoder-decoder architecture. It proposes\revision{ a} structure-induced transformer to capture long-range dependencies and more global information in AST sequences of code snippets.

\textbf{SCRIPT~\cite{2022-SCRIPT}} adopts a Transformer-based encoder-decoder architecture. It proposes two types of Transformer encoders to capture the structural relative positions between tokens for better learning code semantics.

In addition to these non-pre-trained techniques above, since our method is based on the pre-training and fine-tuning paradigm, we also compare two techniques following such \revision{a }paradigm.

\textbf{CodeBERT~\cite{2020-CodeBERT}} is a representative pre-trained model of code. It is trained with the MLM and Replaced Token Detection (RTD) tasks. The authors of CodeBERT fine-tune and test it on the code summarization task (also called the code documentation generation task in their paper). 

\textbf{UniXcoder~\cite{2022-UniXcoder}} is the state-of-the-art pre-trained model of code. It is trained with four tasks: MLM, ULM, Denoising Objective DeNoiSing (DNS), and Code Fragment Representation Learning. Unlike CodeBERT \deletion{that}\revision{which} only pre-trains the encoder, UniXcoder pre-trains both the encoder and the decoder. The authors of UniXcoder also fine-tune and test it on the code summarization task. 

\revision{For non-pre-training-based models (i.e., Re$^2$Com, SiT, and SCRIPT) and pre-training-based models (i.e., CodeBERT and UniXcoder), we train and fine-tune them separately on the training set of each code summarization dataset, and evaluate them on the corresponding test set.}

\noindent2)\textit{\;Results:} \deletion{Table}\revision{TABLE}~\ref{tab:comparison_with_baselines} shows the performances of our {\toolname} and baselines in terms of the three evaluation metrics, i.e., \delete{BLEU-4}\revise{BLEU}, METEOR, and ROUGE-L. \deletion{In Table~\ref{tab:comparison_with_baselines}, $*$ refers to baselines we rerun. The results of the upper part (rows 1--8) are directly reported from~\cite{2021-SiT, 2020-Transformer-based-Approach-for-Code-Summarization, 2021-CAST}. }\revision{In TABLE~\ref{tab:comparison_with_baselines}, }``{\toolname} + CodeBERT'' \deletion{(row 14) }refers to that we build the shared encoder based on the pre-trained encoder provided by CodeBERT\deletion{~\cite{2020-CodeBERT}}. Analogously, in ``{\toolname} + UniXcoder''\deletion{ (row 16)}, the shared encoder and decoder are built upon the pre-trained encoder and decoder provided by UniXcoder\deletion{~\cite{2022-UniXcoder}}. In practice, we initialize our encoder and decoder with the model parameters of the CodeBERT and UniXcoder.\deletion{ {\toolname} (row 12) adopts the same configuration as ``{\toolname} + UniXcoder''.}

\begin{table}[htbp]
    \renewcommand{\arraystretch}{1.2} 
    \caption{Overall performance of baselines and our {\toolname}}
    \label{tab:comparison_with_baselines}
    \scriptsize
    \tabcolsep=2pt
    \centering
    \begin{tabular}{|l|ccc|ccc|}
        \hline
        \multirow{2}{*}{Techniques (Year)} & \multicolumn{3}{c|}{JCSD} & \multicolumn{3}{c|}{PCSD}\\
        
        \cline{2-7}
        
        & \delete{BLEU-4}\revise{BLEU} & METEOR & ROUGE-L & \delete{BLEU-4}\revise{BLEU} & METEOR & ROUGE-L\\
        
        \hline
        
        Re$^2$Com (2020) & 35.65 & 16.26 & 44.95 & -- & -- & --\\
        SiT\revision{ (2021)} & 45.22 & 27.10 & 55.44 & 33.75 & 21.02 & 48.33\\
        SCRIPT (2022) & 46.41 & 28.47 & 56.57 & 33.52 & 20.80 & 48.09 \\
        
        \hline
        \hline
        
        CodeBERT (2020) & 44.80 & 27.73 & 56.00 & 34.35 & 22.02 & 49.69\\ 
        
        {\toolname} + CodeBERT & \textbf{46.21} & \textbf{29.37} & \textbf{56.82} & \textbf{35.34} & \textbf{23.00} & \textbf{50.76}\\
        
        \hline
        \hline
        
        UniXcoder (2022) & 47.15 & 30.02 & 57.98 & 36.03 & 23.37 & 50.93\\ 
        
        {\toolname} + UniXcoder & \textbf{48.31} & \textbf{30.79} & \textbf{58.98} & \textbf{36.36} & \textbf{23.60} & \textbf{51.34}\\
        \hline
    \end{tabular}
\end{table}

From \deletion{rows 1--12 of Table}\revision{TABLE}~\ref{tab:comparison_with_baselines}, it can be observed that, in all non-pre-training baselines, SCRIPT and SiT perform the best on JCSD and PCSD datasets in terms of all three metrics, respectively. However, SCRIPT requires complex preprocessing for code snippets and does not release preprocessing code implementation. Thus, we re-run SCRIPT on their preprocessed datasets. Although the preprocessed datasets are derived from JCSD and PCSD datasets, they have different training and test sets. Therefore, we mainly compare our {\toolname} to SiT in subsequent sections. \deletion{Our }{\toolname}\revision{ built on CodeBERT or UniXcoder} is more powerful than SiT and achieves more impressive performance. On the JCSD dataset, compared with SiT, {\toolname}\revision{ built on UniXcoder} improves by 6.83\% in \delete{BLEU-4}\revise{BLEU}, 13.62\% in METEOR, and 6.39\% in ROUGE-L. On the PCSD dataset, {\toolname}\revision{ built on UniXcoder} also clearly outperforms SiT, improving by 7.73\% in \delete{BLEU-4}\revise{BLEU}, 12.27\% in METEOR, and 6.23\% in ROUGE-L. \revision{Because {\toolname} built upon UniXcoder performs the best, unless explicitly stated, {\toolname} appearing alone refers to ``{\toolname} + UniXcoder'' by default.}

\begin{figure*}[!t]
\centering
\subfigure[BLEU Score Distribution]
    {
        \includegraphics[width=0.3\linewidth]{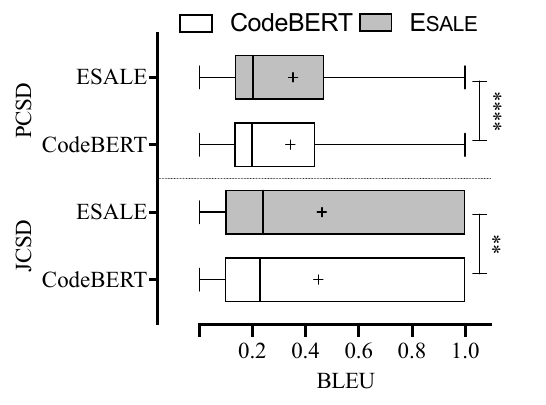}
        \label{fig:BLEU-ESALE_vs_CodeBERT}
    }
    \subfigure[METEOR Score Distribution]
    {
        \includegraphics[width=0.3\linewidth]{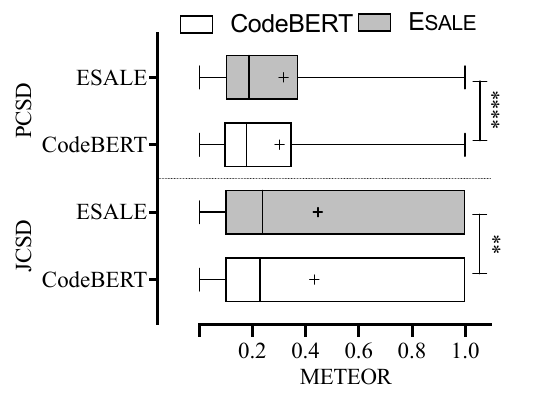}
        \label{fig:METEOR-ESALE_vs_CodeBERT}
    }
    \subfigure[ROUGE-L Score Distribution]
    {
        \includegraphics[width=0.3\linewidth]{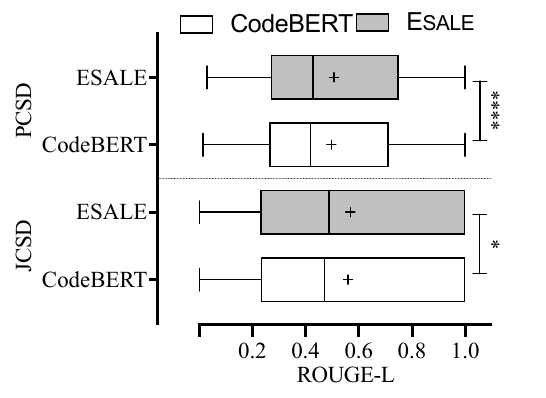}
        \label{fig:ROUGE-L-ESALE_vs_CodeBERT}
    }
    \subfigure[BLEU Score Distribution]
    {
        \includegraphics[width=0.3\linewidth]{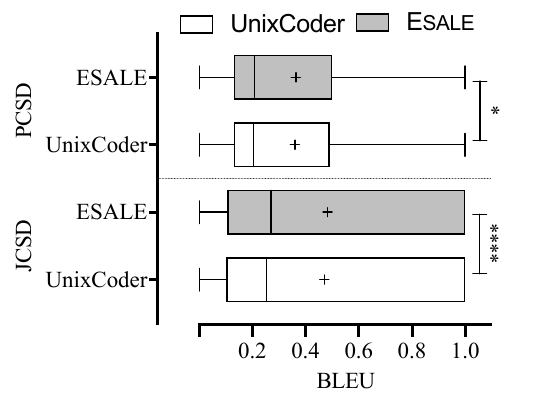}
        \label{fig:BLEU-ESALE_vs_UniXcoder}
    }
    \subfigure[METEOR Score Distribution]
    {
        \includegraphics[width=0.3\linewidth]{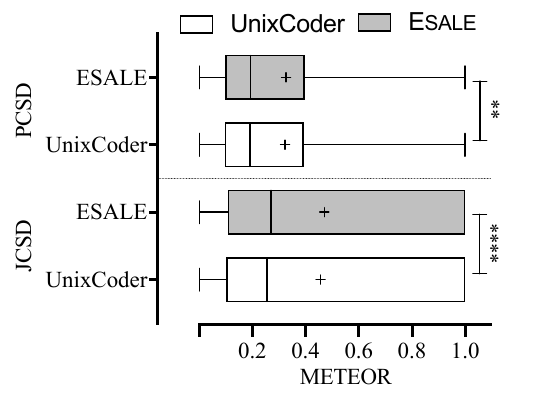}
        \label{fig:METEOR-ESALE_vs_UniXcoder}
    }
    \subfigure[ROUGE-L Score Distribution]
    {
        \includegraphics[width=0.3\linewidth]{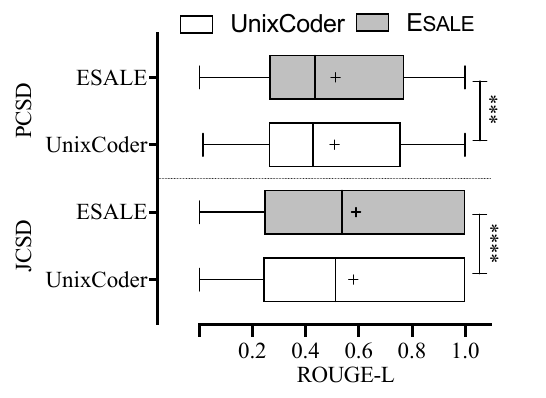}
        \label{fig:ROUGE-L-ESALE_vs_UniXcoder}
    }
    \caption{Score distribution of three metrics. ``*'' ($0.01 < p < 0.05$), ``**" ($0.001 < p < 0.01$), ``***'' ($0.0001 < p < 0.001$) and ``****'' ($p < 0.0001$) represent the differences between two groups are Significant, Very significant, Extremely significant and Extremely significant, respectively. And ‘ns’ ($p \geq 0.05$) means Not significant.}
    \label{fig:distribution_of_metrics}
\end{figure*}

\deletion{As mentioned earlier, we build our model upon the pre-trained models, such as CodeBER\revision{T} and UniXcoder. These pre-trained models are also evaluated on the code summarization tasks in their paper. Thus, we re-run and compare two pre-trained models on the JCSD and PCSD datasets, and the experimental results are shown in \delete{the }rows 13--16 of \deletion{Table}\revision{TABLE}~\ref{tab:comparison_with_baselines}. From these rows,}\revision{In addition,} it can be observed that, our method consistently improves the performance of the original pre-trained models\revision{, i.e., CodeBERT and UniXcoder} on both datasets in general. It should be noted that the values in \deletion{Table}\revision{TABLE}~\ref{tab:comparison_with_baselines} are the average scores of all test samples. For a more comprehensive comparison, we further compare the distribution of the scores of CodeBERT, UniXcoder and \deletion{our }{\toolname} on all test samples, and the statistical results are shown in \deletion{Figure}\revision{Fig.}~\ref{fig:distribution_of_metrics}. In \deletion{Figure}\revision{Fig.}~\ref{fig:distribution_of_metrics}, `+' denotes the mean, which is the value filled in \deletion{Table}\revision{TABLE}~\ref{tab:comparison_with_baselines}. Overall, the score distribution of {\toolname} is better than that of pre-trained models (i.e., CodeBERT and \delete{UnixCoder}\revise{UniXcoder}). To test whether there is a statistically significant difference between {\toolname} and pre-trained models, we perform the paired Wilcoxon-Mann-Whitney signed-rank test at a significance level of 5\%, following previously reported guidelines for inferential statistical analysis involving randomized algorithms~\cite{2014-Hitchhiker-Guide-Statistical-Tests, 2015-Practical-Test-Selection}. From \deletion{Figure}\revision{Fig.}~\ref{fig:distribution_of_metrics}, it can be observed that, intuitively, in all three metrics, {\toolname} outperforms pre-trained models CodeBERT and UniXcoder on both the JCSD and PCSD datasets. In summary, the results and observations above demonstrate that under all experimental settings, our {\toolname} consistently achieves higher performance in all three metrics, which indicates better code summarization performance. Note that our {\toolname} is non-intrusive, indicating that it can be combined with future state-of-the-art language models to further improve code summarization performance.

\begin{table}[htbp]
    \centering
    \tabcolsep=2pt
    \footnotesize
    \renewcommand{\arraystretch}{1.2}
    \caption{\revise{Effectiveness of {\toolname} on the deduplicated JCSD and PCSD datasets}}
    \label{tab:comparison_with_baselines_on_deduplicate_JCSD_PCSD}
    \begin{tabular}{|l|ccc|ccc|}
        \hline
        \multirow{2}{*}{\revise{Technique}} & \multicolumn{3}{c|}{\revise{JCSD}} & \multicolumn{3}{c|}{\revise{PCSD}}\\
        
        \cline{2-7}
        
        & \revise{BLEU} & \revise{METEOR} & \revise{ROUGE-L} & \revise{BLEU} & \revise{METEOR} & \revise{ROUGE-L}\\
        
        \hline
    
        \revise{SiT} & \revise{27.98} & \revise{16.38} & \revise{41.15} & \revise{33.76} & \revise{21.03} & \revise{48.35} \\
    
        \revise{SCRIPT} & \revise{29.24} & \revise{17.74} & \revise{42.44} & \revise{33.51} & \revise{20.80} & \revise{48.09} \\
        
        \revise{CodeBERT} & \revise{44.64} & \revise{27.60} & \revise{55.86} & \revise{34.33} & \revise{22.02} & \revise{49.68}\\ 
        
        \revise{UniXcoder} & \revise{46.85} & \revise{29.89} & \revise{57.85} & \revise{35.99} & \revise{23.33} & \revise{50.90} \\ 
    
        \hline
        \revise{{\toolname}} & \revise{\textbf{48.05}} & \revise{\textbf{30.69}} & \revise{\textbf{58.89}} & \revise{\textbf{36.34}} & \revise{\textbf{23.61}} & \revise{\textbf{51.33}} \\
        
        \hline
    \end{tabular}
\end{table}

\begin{table}[htbp]
    \centering
    \tabcolsep=2pt
    \footnotesize
    \renewcommand{\arraystretch}{1.2}
    \caption{\revision{Effectiveness of {\toolname} on the JCSD and PCSD datasets from~\cite{2022-Data-Preprocessing-for-Code-Summarization}}}
    \label{tab:comparison_with_baselines_on_filtered_JCSD_PCSD}
    \begin{tabular}{|l|ccc|ccc|}
        \hline
        \multirow{2}{*}{\revision{Technique}} & \multicolumn{3}{c|}{\revision{JCSD}} & \multicolumn{3}{c|}{\revision{PCSD}}\\
        
        \cline{2-7}
        
        & \revision{BLEU} & \revision{METEOR} & \revision{ROUGE-L} & \revision{BLEU} & \revision{METEOR} & \revision{ROUGE-L}\\
        
        \hline
    
        \revision{SiT} & \revision{11.24} & \revision{9.86} & \revision{24.30} & \revision{15.40} & \revision{9.56} & \revision{23.08}\\ 
        
        \revision{CodeBERT} & \revision{27.88} & \revision{19.00} & \revision{39.05} & \revision{16.14} & \revision{11.37} & \revision{25.23}\\ 
        
        \revision{UniXcoder} & \revision{29.00} & \revision{20.36} & \revision{40.31} & \revision{17.10} & \revision{12.19} & \revision{26.80}\\  
    
        \hline
        \revision{{\toolname}} & \revision{\textbf{30.52}} & \revision{\textbf{21.63}} & \revision{\textbf{42.63}} & \revision{\textbf{17.94}} & \revision{\textbf{12.82}} & \revision{\textbf{28.14}} \\
        
        \hline
    \end{tabular}
\end{table}

\begin{table}[htbp]
    \renewcommand{\arraystretch}{1.2}  
    \footnotesize
    \tabcolsep=4pt
    \caption{\revise{Effectiveness of {\toolname} on the CPJD dataset}}
    \label{tab:comparison_with_baselines_on_CPJD}
    \centering
    \begin{tabular}{|l|ccc|}
        \hline
        \multirow{2}{*}{\revise{Technique}} & \multicolumn{3}{c|}{\revise{CPJD}} \\
        
        \cline{2-4}
        
        & \revise{BLEU} & \revise{METEOR} & \revise{ROUGE-L} \\
        
        \hline
        
        \revise{SiT} & \revise{10.35} & \revise{6.75} & \revise{14.06} \\
        
        \revise{CodeBERT} & \revise{15.40} & \revise{10.05} & \revise{19.29} \\ 
        
        \revise{UniXcoder} & \revise{20.62} & \revise{13.11} & \revise{24.44} \\ 

        \hline

        \revise{{\toolname}} & \revise{\textbf{21.42}} & \revise{\textbf{13.79}} & \revise{\textbf{25.94}} \\
        
        \hline
    \end{tabular}
\end{table}

\revise{\deletion{As mentioned in Section 5.1.1, existing}\revision{Existing} research~\cite{2022-Evaluation-Neural-Code-Summarization} find that there are different degrees of data overlap in the training sets and test sets of \revision{the original }JCSD and PCSD\revision{ datasets}, respectively. Considering that data leakage may affect the accuracy of performance evaluation, we further remove the duplicate code snippets in the training and test sets of these two datasets, and measure the performance of {\toolname} and several baselines again. TABLE~\ref{tab:comparison_with_baselines_on_deduplicate_JCSD_PCSD} presents their performance on the deduplicated JCSD and PCSD datasets. \deletion{From this table, it}\revision{It} is observed that {\toolname} consistently outperforms the four baselines\deletion{ (including SiT, SCRIPT, CodeBERT, and UniXcoder) in terms of}\revision{ in} all three metrics.}

\revision{Shi et al.~\cite{2022-Data-Preprocessing-for-Code-Summarization} find that there are some noisy data in the JCSD and PCSD datasets and provide filtered versions of these two datasets after removing the noisy data. Therefore, we also compare {\toolname} with the three baselines (SiT, CodeBERT, and UniXcoder) on the JCSD and PCSD datasets released by Shi et al.~\cite{2022-Data-Preprocessing-for-Code-Summarization}. SCRIPT requires special data preprocessing, and we failed to reproduce it on these two filtered datasets. The results are shown in TABLE~\ref{tab:comparison_with_baselines_on_filtered_JCSD_PCSD}, demonstrating that {\toolname} outperforms the best baseline UniXcoder in all metrics.}

\begin{table*}[htbp]
    \scriptsize
    \centering
    \renewcommand{\arraystretch}{1.2}
    \tabcolsep=4pt
    \caption{\revise{Effectiveness of {\toolname} on other programming language datasets, including CSN-PHP, -GO, -JavaScript, and -Ruby.}}
    \label{tab:compare_baseline_on_CodeXGLEU}
    \begin{tabular}{|l|ccc|ccc|ccc|ccc|}
    
        \hline
        \multirow{2}{*}{\revise{Technique}} &
        \multicolumn{3}{c|}{\revise{CSN-PHP}} & \multicolumn{3}{c|}{\revise{CSN-Go}} &
        \multicolumn{3}{c|}{\revise{CSN-JavaScript}} & \multicolumn{3}{c|}{\revise{CSN-Ruby}} \\
        
        \cline{2-13}
        
        & \revise{BLEU} & \revise{METEOR} & \revise{ROUGE-L} & \revise{BLEU} & \revise{METEOR} & \revise{ROUGE-L} & \revise{BLEU} & \revise{METEOR} & \multicolumn{1}{c|}{\revise{ROUGE-L}} & \revise{BLEU} & \revise{METEOR} & \multicolumn{1}{c|}{\revise{ROUGE-L}} \\
        
        \hline
        
        \revise{CodeBERT} & \revise{23.11} & \revise{15.16} & \revise{38.69} & \revise{16.59} & \revise{10.47} & \revise{31.06} & \revise{14.06} & \revise{9.30} & \revise{25.97} & \revise{12.70} & \revise{8.94} & \multicolumn{1}{c|}{\revise{20.11}} \\ 
        
        \revise{UniXcoder} & \revise{26.36} & \revise{16.53} & \revise{41.75} & \revise{17.78} & \revise{11.69} & \revise{33.88} & \revise{15.46} & \revise{9.74} & \revise{26.77} & \revise{14.85} & \revise{9.88} & \multicolumn{1}{c|}{\revise{26.57}} \\

        \hline

        \revise{{\toolname}} & \revise{\textbf{26.76}} & \revise{\textbf{16.62}} & \revise{\textbf{41.84}} & \revise{\textbf{18.15}} & \revise{\textbf{11.74}} & \revise{\textbf{34.07}} & \revise{\textbf{15.61}} & \revise{\textbf{9.83}} & \revise{\textbf{26.90}} & \revise{\textbf{14.99}} & \revise{\textbf{9.89}} & \multicolumn{1}{c|}{\revise{\textbf{26.68}}}\\
   
        \hline
    \end{tabular}
\end{table*}

\revise{As mentioned earlier, we also conduct an experiment on a cross-project dataset called CPJD. TABLE~\ref{tab:comparison_with_baselines_on_CPJD} shows the overall performance of the \deletion{four}\revision{three} baselines\deletion{ (including SiT, CodeBERT, and UniXcoder)} and our {\toolname} on the \deletion{CPD}\revision{CPJD} dataset. SCRIPT requires special data preprocessing, and we failed to reproduce it on CPJD. From TABLE~\ref{tab:comparison_with_baselines_on_CPJD}, it is observed that our {\toolname} outperforms all three baselines in terms of all three metrics.
}

\revise{Moreover, we conduct experiments on the CSN dataset to evaluate the effectiveness and generalizability of {\toolname} on more programming languages. TABLE~\ref{tab:compare_baseline_on_CodeXGLEU} shows the overall performance of CodeBERT, UniXcoder, and our {\toolname} on the CSN-PHP, -Go, -JavaScript, and -Ruby datasets. The baselines SiT and SCRIPT require special data preprocessing, which cannot be applied to some programming languages in the CSN dataset (e.g., Go, PHP, and Ruby). Therefore, on the CSN dataset, we only \deletion{compared}\revision{compare {\toolname} with} CodeBERT and UniXcoder. From TABLE~\ref{tab:compare_baseline_on_CodeXGLEU}, it is observed that {\toolname} consistently outperforms CodeBERT and UniXcoder on the four programming language code summarization tasks.}

\begin{table}[htbp]
    \centering
    \tabcolsep=1.5pt
    \scriptsize
    \renewcommand{\arraystretch}{1.2}
    \caption{\revision{Performance of {\toolname} when the pre-trained shared encoder is frozen during fine-tuning on downstream code summarization tasks}}
    \label{tab:effectiveness_when_froze_encoder}
    \begin{tabular}{|l|ccc|ccc|}
        \hline
        \multirow{2}{*}{\revision{Technique}} & \multicolumn{3}{c|}{\revision{JCSD}} & \multicolumn{3}{c|}{\revision{PCSD}}\\
        
        \cline{2-7}
        
        & \revision{BLEU} & \revision{METEOR} & \revision{ROUGE-L} & \revision{BLEU} & \revision{METEOR} & \revision{ROUGE-L}\\
        
        \hline
        
        \revision{CodeBERT$_{frozen}$} & \revision{18.59} & \revision{11.13} & \revision{30.67} & \revision{18.29} & \revision{9.18} & \revision{29.62}\\ 
        
        \revision{UniXcoder$_{frozen}$} & \revision{20.25} & \revision{12.38} & \revision{31.49} & \revision{18.62} & \revision{10.04} & \revision{30.01}\\  
    
        \hline
        \revision{{\toolname}$_{frozen}$} & \revision{\textbf{25.15}} & \revision{\textbf{15.31}} & \revision{\textbf{35.57}} & \revision{\textbf{21.52}} & \revision{\textbf{13.62}} & \revision{\textbf{34.18}} \\
        
        \hline
    \end{tabular}
    \vspace{-4mm}
\end{table}

\revision{To accurately reflect whether the performance improvement is attributable to the shared encoder trained with three summary-focused pre-training tasks, we further conduct experiments where the pre-trained encoder remains frozen while only the decoder undergoes fine-tuning. Rows CodeBERT$_{frozen}$, UniXcoder$_{frozen}$, and {\toolname}$_{frozen}$ of TABLE~\ref{tab:effectiveness_when_froze_encoder} show the performance of baselines and {\toolname} when the parameters of the pre-trained shared encoder are frozen during fine-tuning on downstream code summarization tasks. It is observed that compared to code summarization models built on unfrozen encoders, CodeBERT$_{frozen}$, UniXcoder$_{frozen}$, and {\toolname}$_{frozen}$ employing frozen encoders all exhibit varying degrees of performance degradation. It is reasonable because freezing the parameters of the pre-trained encoder can restrict the model's adaptability and hinder its ability to effectively learn task-specific features. Furthermore, it can be observed that {\toolname}$_{frozen}$ outperforms CodeBERT$_{frozen}$ and UniXcoder$_{frozen}$ in all metrics, which effectively demonstrates the capability of the three summary-focused pre-training tasks to enhance the encoder's code summarization ability.
}

\subsubsection{\textbf{RQ2:} Effect of Each Pre-train Tasks (Ablation Study)}
\label{subsubsec:results_of_RQ2}

We use three tasks (AWP, MLM, and ULM) to enhance the ability of our model to learn code-summary alignment. We conduct ablation studies to reveal the influence of each task on the performance of {\toolname}. \deletion{And the}\revision{The} study results are shown in \deletion{Table}\revision{TABLE}~\ref{tab:ablation_study}, in which ``{\toolname}\delete{-}\revise{ w/o }AWP''\deletion{ (Row 4)}, ``{\toolname}\delete{-}\revise{ w/o }MLM''\deletion{ (Row 5)}, and ``{\toolname}\delete{-}\revise{ w/o }ULM''\deletion{ (Row 6)} mean that we train {\toolname} without the AWP, MLM, and ULM tasks, respectively\deletion{, and ``{\toolname}\delete{-AWP-MLM-UML}\revise{ w/o AWP, MLM, ULM}'' (Row \deletion{7}\revision{5}) means that we train {\toolname} without the AWP, MLM, and ULM tasks (i.e., UniXcoder)}. \deletion{It can be observed}\revision{It is observed} that the performance of {\toolname} degrades when any of the three tasks are ignored. Therefore, it can be concluded that all \deletion{the }three tasks play an important role in improving the code summarization performance of {\toolname}. 

\begin{figure}[htbp]
  \centering
  \includegraphics[width=\linewidth]{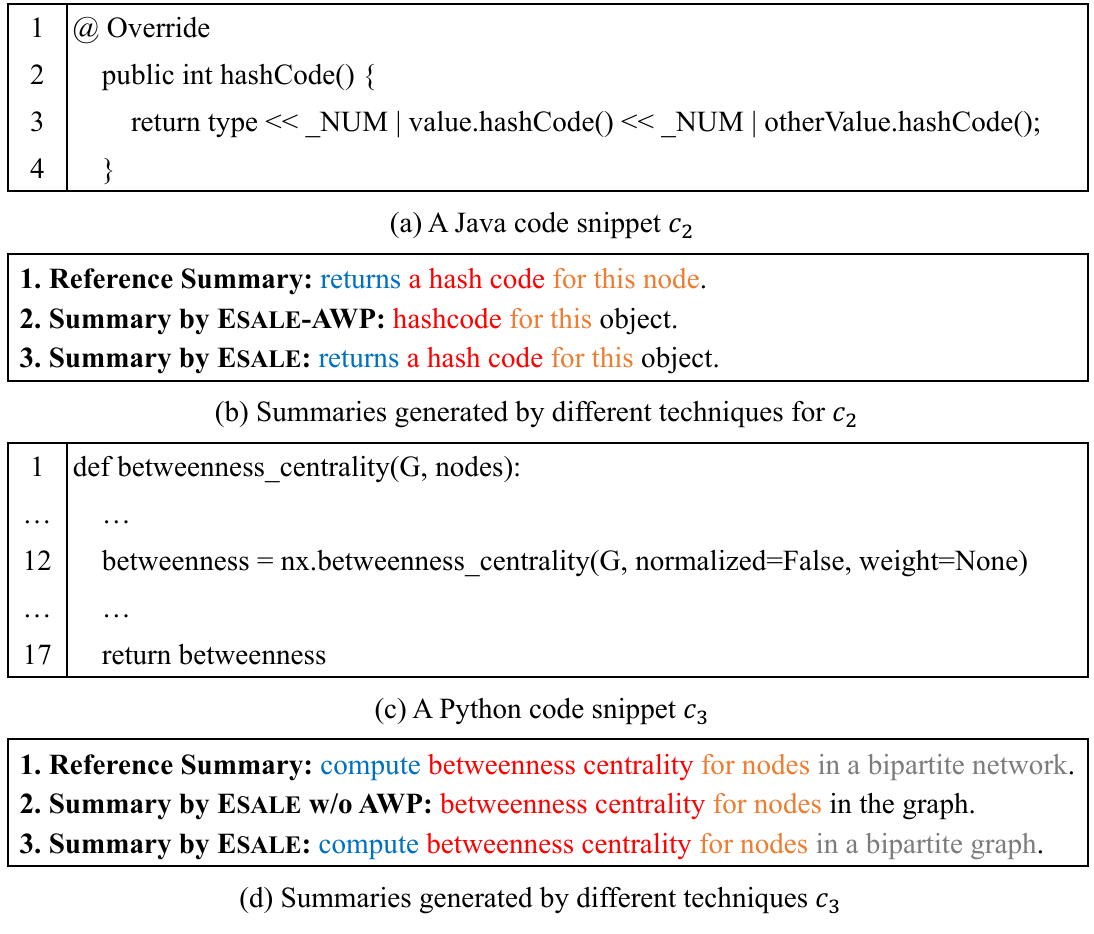}
  \vspace{-4mm}
  \caption{Examples of AWP contributions}
  \label{fig:example_of_AWP_advantages}
\end{figure}

In addition, from\deletion{ \delete{the }rows \deletion{4-6}\revision{2-4} of} \deletion{Table}\revision{TABLE}~\ref{tab:ablation_study}, it can be observed that the AWP task has the most significant effect on the performance of {\toolname}. We further delve into the contribution of the AWP task to {\toolname}, which is a task especially designed for the code summarization~\cite{2021-Action-Word-Prediction-for-Code-Summarization}. 
\deletion{As mentioned in Section~\ref{subsubsec:shared_encoder_training}(i), we follow the work~\cite{2021-Action-Word-Prediction-for-Code-Summarization} and use the top-40 setting, that is, the model attempts to predict the forty most-common action words, or ``other'' if predicted to be a less-common action word. }In \deletion{Table}\revision{TABLE}~\ref{tab:samples_with_AW_in_top40}, the second column shows the total number of samples in the JCSD and PCSD test set, and the ``Num.'' and ``Pro.'' columns show the number and proportion of samples whose action words are included in the top-40 common list, respectively. From this table, it can be observed that the top-40 setting only covers about 61\% of samples in test sets. In other words, many action words (in the remaining 39\% samples) are still not included. We further compute the \deletion{action word prediction}\revision{AWP} accuracy of {\toolname}, and the results are shown in the ``AWP Acc.'' column of \deletion{Table}\revision{TABLE}~\ref{tab:samples_with_AW_in_top40}. It can be observed that the average \deletion{action word prediction}\revision{AWP} accuracy of {\toolname} is 64.89\%. We also check whether the performance of {\toolname} can be improved when the action words are correctly predicted. \deletion{Table}\revision{TABLE}~\ref{tab:correct_AWP_contribution} shows the results of {\toolname} and {\toolname}\delete{-}\revise{ w/o} AWP on the samples whose action words are included in the top 40 common list and predicted correctly. In \deletion{Table}\revision{TABLE}~\ref{tab:correct_AWP_contribution}, ``{\toolname}\delete{-}\revise{ w/o }AWP'' means that we train {\toolname} without the AWP task; ``\# Improved'' denotes the number of the samples for which \delete{A}\revise{{\toolname}} can generate higher quality summaries (i.e., larger BLEU-4, METEOR, and ROUGE-L) when correctly predicting their action words. From \deletion{Table}\revision{TABLE}~\ref{tab:correct_AWP_contribution}, it can be observed that {\toolname} can generate higher quality summaries for 914 Java and 2,283 Python code snippets on average when predicting their action words correctly. For each metric, we also perform the paired Wilcoxon-Mann-Whitney signed-rank test on all scores got by {\toolname}\delete{-}\revise{ w/o} AWP and {\toolname} at a significance level of 5\%. The test results are presented in the ``$p$-value'' columns of \deletion{Table}\revision{TABLE}~\ref{tab:correct_AWP_contribution}. We can intuitively observe that in all three metrics, {\toolname} outperforms {\toolname}\delete{-}\revise{ w/o} AWP on both the JCSD and PCSD datasets, which means the AWP plays a significant role in facilitating {\toolname} to generate high-quality summaries, as claimed by~\cite{2021-Action-Word-Prediction-for-Code-Summarization}. \deletion{Figure}\revision{Fig.}~\ref{fig:example_of_AWP_advantages} shows two examples, including a Java example \delete{$c_3$}\revise{$c_2$} and a Python example \delete{$c_4$}\revise{$c_3$}. From these two examples, we can also intuitively observe that compared to {\toolname}\delete{-}\revise{ w/o }AWP, {\toolname} can generate higher quality and closer reference summaries, which can be attributed to correctly predicted action words.

\begin{table}[htbp]
    \centering
    \renewcommand{\arraystretch}{1.2} 
    \scriptsize
    \tabcolsep=2pt
    \caption{Effect of three pre-training tasks AWP, MLM, and ULM\deletion{\revise{. {\toolname} \delete{-}\revise{w/o} AWP, \deletion{{\toolname} w/o }MLM, and \deletion{{\toolname} w/o }ULM denote that we pre-train the shared encoder of {\toolname} without AWP, MLM, and ULM, respectively.
    }}
    }
    \label{tab:ablation_study}
  
    \begin{tabular}{|l|ccc|ccc|}
        \hline
        \multirow{2}{*}{Technique\delete{s}} & \multicolumn{3}{c|}{JCSD} & \multicolumn{3}{c|}{PCSD}\\
        \cline{2-7}
        & \delete{BLEU-4}\revise{BLEU} & METEOR & ROUGE-L & BLEU & METEOR & ROUGE-L\\
        
        \hline
        
        {\toolname} \delete{-}\revise{w/o} AWP & 47.86 & 30.54 & 58.55 & 36.10 & 23.39 & 51.02\\
        
        {\toolname} \delete{-}\revise{w/o} MLM & 48.13 & 30.6 & 58.67 & 36.25 & 23.45 & 51.18\\
        
        {\toolname} \delete{-}\revise{w/o} ULM & 47.92 & 30.54 & 58.65 & 36.14 & 23.47 & 51.10\\
        
        \hline
    \end{tabular}
\end{table}

\begin{table}[htbp]
  \renewcommand{\arraystretch}{1.2}
  \caption{The statistic information of samples in test sets whose action words (AW) are included in \revision{the} pre-defined top 40 common list. ``Num.", ``Pro.'', and ``Acc.'' are the abbreviations of the words ``Number'', ``Proportion'', and ``Accuracy", respectively. }
  \footnotesize
  \label{tab:samples_with_AW_in_top40}
  \centering
  \begin{tabular}{|c|c|c|c|c|}
    \hline
    \multirow{2}{*}{Dataset} & \multirow{2}{*}{Test Set} & \multicolumn{2}{c|}{With Common AW} & \multirow{2}{*}{AWP Acc.}\\
    \cline{3-4}
    & & Num. & Pro. & \\
    \hline
    JCSD & 8,714 & 5,351 & 61.41\% & 72.42\%\\
    PCSD & 19,066 & 11,583 & 60.75\% & 57.35\%\\
    \hline
    \hline
    \multicolumn{3}{|c|}{Average} & 61.08\% & 64.89\% \\
    \hline
 \end{tabular}
\end{table}

\begin{table*}[htbp]
    \centering
    \renewcommand{\arraystretch}{1.2}
    \footnotesize
    \caption{Results of {\toolname} and {\toolname} \delete{-}\revise{w/o} AWP on the samples whose action words are included in \revision{the }top 40 and predicted correctly. In the ``$p$-value'' columns, the symbol ``*'' has the same meaning as that in Figure~\ref{fig:distribution_of_metrics}. \revise{\# Improved denotes the number of the samples for which {\toolname} can generate higher quality summaries when correctly predicting their action words.}}
    \label{tab:correct_AWP_contribution}
    \begin{tabular}{|l|c|c|c|c|c|c|c|c|}
        \hline
        \multirow{2}{*}{Metric\delete{s}} & \multicolumn{4}{c|}{JCSD} & \multicolumn{4}{c|}{PCSD}\\
        \cline{2-9}
        
        & \# Improved & {\toolname} \delete{-}\revise{w/o} AWP & {\toolname} & $p$-value & \# Improved & {\toolname} \delete{-}\revise{w/o} AWP & {\toolname} & $p$-value\\
        
        \hline
        
        \delete{BLEU-4}\revise{BLEU} & 890 & 21.53 & 35.70 & **** & 2,236 & 23.77 & 39.66 & **** \\
        
        \hline
        
        METEOR & 940 & 20.18 & 33.97 & **** & 2,327 & 20.75 & 36.27 & **** \\
        
        \hline
        
        ROUGE-L & 913 & 39.52 & 55.52 & **** & 2,286 & 44.66 & 62.56 & **** \\
        
        \hline
        \hline
        
        Average & 914 & 27.08 & 41.73 & \revise{****} & 2,283 & 29.73 & 46.16 & \revise{****} \\
        
        \hline
    \end{tabular}
\end{table*}

\subsubsection{\textbf{RQ3:} Robustness of {\toolname}}
\label{subsubsec:robustness_of_our_approach} 
\begin{figure*}[htbp]
\centering
    \subfigure[Code in JCSD Test Set]
    {
        \includegraphics[width=0.22\linewidth]{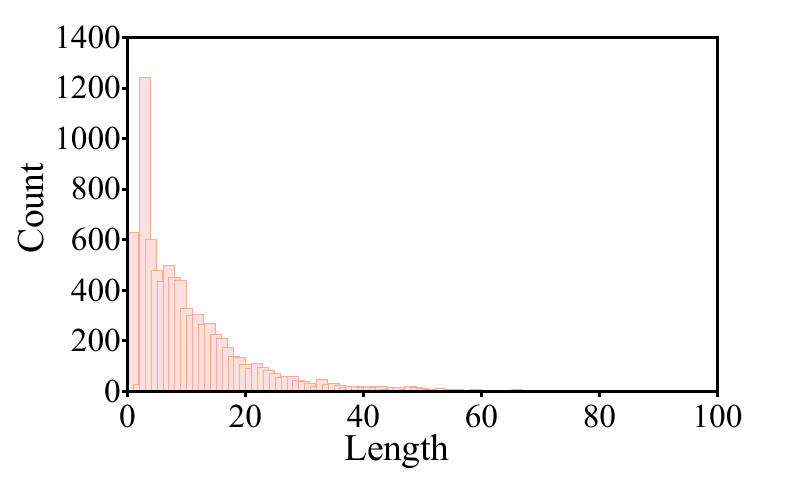}
        \label{fig:distribution_of_code_length_JCSD_test_set}
    }
    \subfigure[Comments in JCSD Test Set]
    {
        \includegraphics[width=0.22\linewidth]{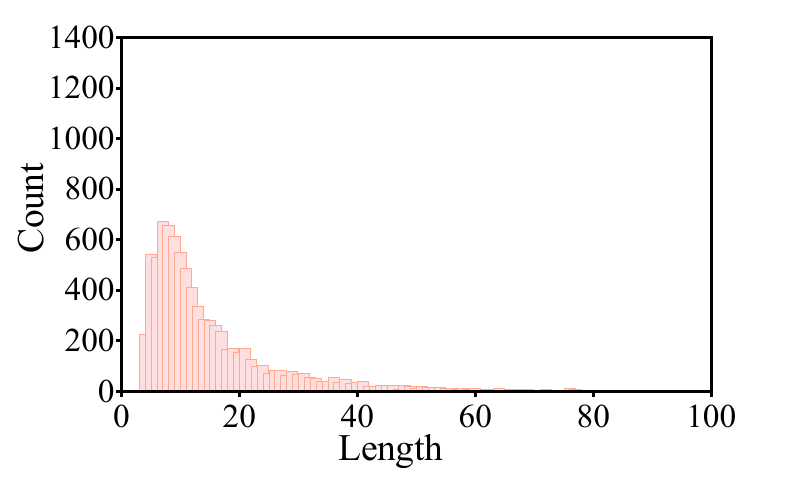}
        \label{fig:distribution_of_comment_length_JCSD_test_set}
    }
    \subfigure[Code in PCSD Test Set]
    {
        \includegraphics[width=0.22\linewidth]{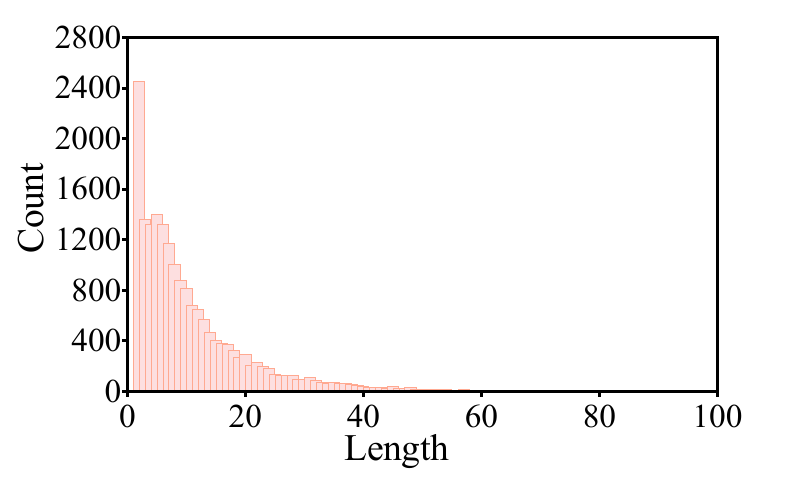}
        \label{fig:distribution_of_code_length_PCSD_test_set}
    }
    \subfigure[Comments in PCSD Test Set]
    {
        \includegraphics[width=0.22\linewidth]{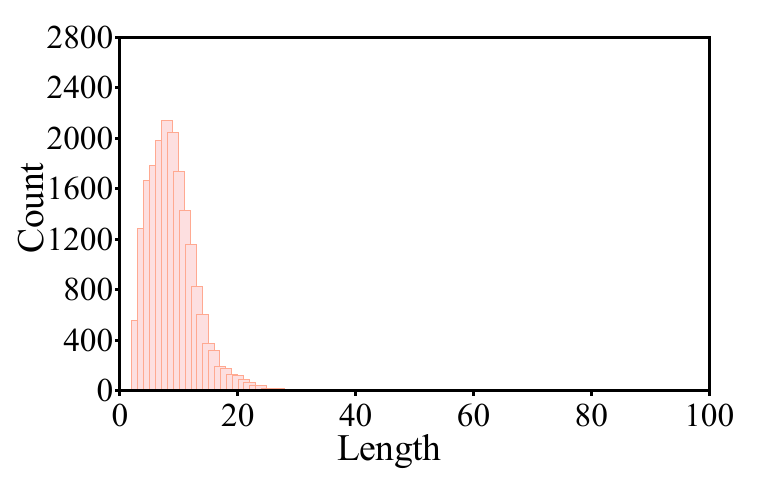}
        \label{fig:distribution_of_comment_length_PCSD_test_set}
    }
    \caption{Length distribution of code snippets and comments in test sets}
    \label{fig:distribution_of_length}
    \vspace{-2mm}
\end{figure*}

To analyze the robustness of {\toolname}, we study two parameters (i.e., code length and comment length) that may \deletion{have an influence on}\revision{influence} the embedding representations of code snippets and comments. 
\deletion{Figure}\revision{Fig.}~\ref{fig:distribution_of_length} shows the length distributions of code snippets and comments on the test sets of the JCSD and PCSD datasets. For a code snippet, its length refers to the lines of the code snippet. For a comment, its length refers to the number of words in the comment. From \deletion{Figure}\revision{Fig.}~\ref{fig:distribution_of_length} (a) and (c), it can be observed that most code snippets are between 20 and 40 lines. From \deletion{Figure}\revision{Fig.} \ref{fig:distribution_of_length} (b) and (d), it is noticed that almost all comments are less than 20 in length. This also confirms the challenge of capturing the correlation between the long code snippet \deletion{with}\revision{and} its \deletion{corresponding }short comment (summary).

\begin{figure*}[htbp]
\centering
    \subfigure[Code in JCSD Test Set]
    {
        \includegraphics[width=0.22\linewidth]{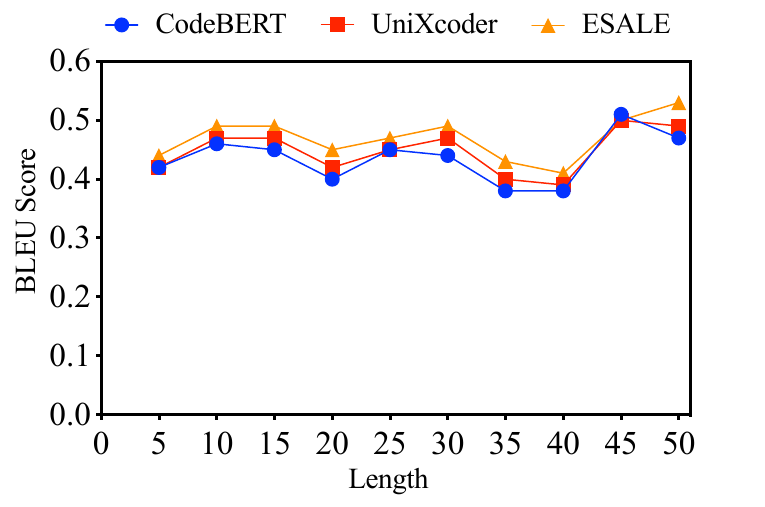}
        \label{fig:Robustness_on_vary_code_length_JCSD_test_set}
    }
    \subfigure[Comments in JCSD Test Set]
    {
        \includegraphics[width=0.22\linewidth]{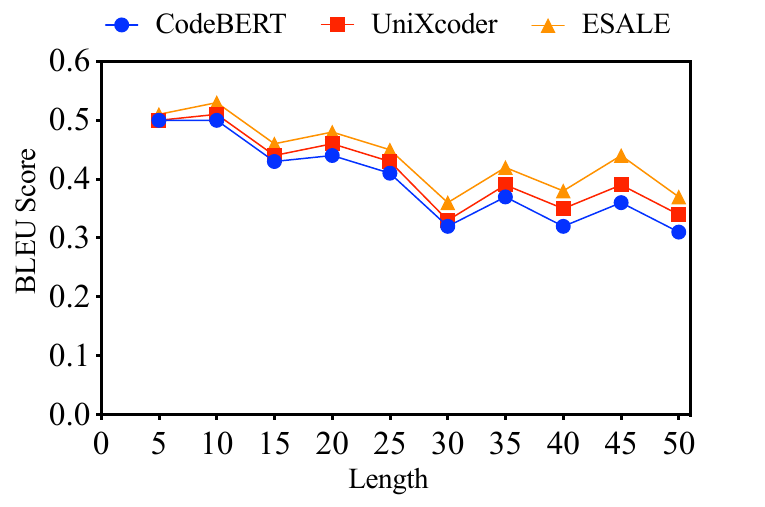}
        \label{fig:Robustness_on_vary_comment_length_JCSD_test_set}
    }
    \subfigure[Code in PCSD Test Set]
    {
        \includegraphics[width=0.22\linewidth]{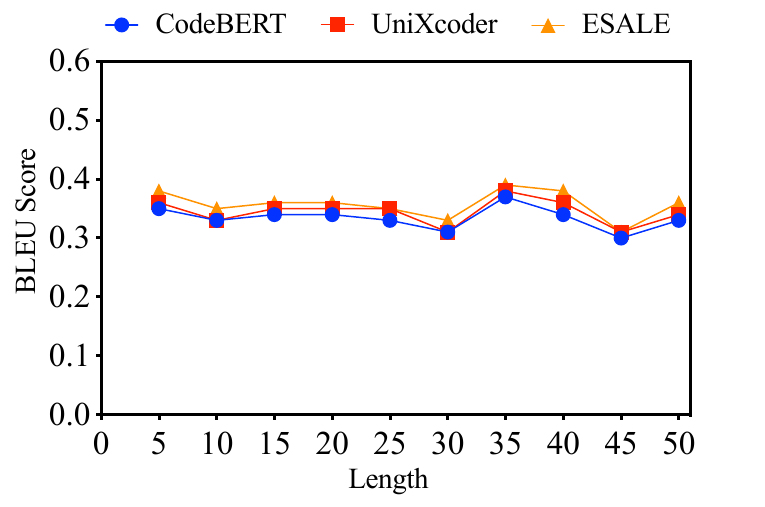}
        \label{fig:Robustness_on_vary_code_length_PCSD_test_set}
    }
    \subfigure[Comments in PCSD Test Set]
    {
        \includegraphics[width=0.22\linewidth]{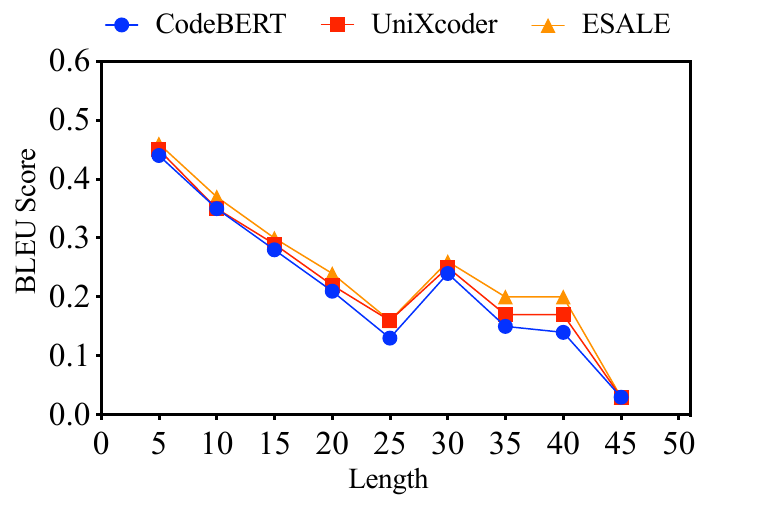}
        \label{fig:Robustness_on_vary_comment_length_PCSD_test_set}
    }
    \caption{Effect of code snippet and comment length on the robustness of {\toolname}\revision{. The values shown in the figures are averaged results across length intervals (e.g., [0-5], [6-10], \dots, [46-50]) rather than specific lengths.}}
    \label{fig:robustness_of_EACS}
    \vspace{-2mm}
\end{figure*}

\deletion{Figure}\revision{Fig.}~\ref{fig:robustness_of_EACS} shows the performance of \revision{two best baselines (i.e., CodeBERT and UniXcoder) and }{\toolname} based on \deletion{different evaluation metrics}\revision{the BLEU metric} with varying parameters. \revision{In this figure, the version of {\toolname} is the same as ``{\toolname} + UniXcoder''.} From \deletion{Figure}\revision{Fig.}~\ref{fig:robustness_of_EACS} (a) and (b), it can be observed that on the JCSD test set, {\toolname} maintains stable performance even though the code snippet length or comment length increases. On the PCSD test set, from \deletion{Figure}\revision{Fig.}~\ref{fig:robustness_of_EACS}(c) and (d), it can be observed that {\toolname} also maintains stable performance when the code snippet length increases, however, the performance of {\toolname} degrades significantly as the length of the comment increases.\deletion{ We further investigate the performance of SiT and UniXcoder on varying the PCSD comments, and the results are shown in \deletion{Figure}\revision{Fig.}~\ref{fig:robustness_of_baselines}. \revision{We run SiT and UniXcoder on the same PCSD test set as {\toolname}.  
SiT and UniXcoder as state-of-the-art code summarization techniques representing distinct paradigms: SiT is a non-pre-trained-based technique, while UniXcoder is a pre-trained-based technique. }From this figure, we can observe the same phenomenon as {\toolname}. }
\revision{In addition, it is observed that the same phenomenon occurred in CodeBERT and UniXcoder.}
It means that as the expected length of the generated summary continues to increase, it will be more challenging to generate. Overall, the results above verify the robustness of our {\toolname}. \revision{Due to the page limit, please refer to our project website~\cite{2024-ESALE-google-site} for additional results on the METEOR and ROUGE-L metrics, where you can find that the robustness of both the baselines and {\toolname} on METEOR and ROUGE-L is similar to that on BLEU.}

\begin{table}[t]
    \renewcommand{\arraystretch}{1.2}
    \footnotesize
    \tabcolsep=2.5pt
    \caption{Results of human evaluation. The first and second values in parentheses represent standard deviation and significant difference, respectively. The symbol ``*'' has the same meaning as that in Figure~\ref{fig:distribution_of_metrics}.}
    \label{tab:human_evaluation}
    \centering
    \begin{tabular}{|c|l|l|l|l|}
    \hline
    Dataset & Metrics & SiT & UniXcoder & {\toolname}\\
    \hline
    \multirow{4}{*}{JCSD} & Similarity & 2.06 (0.51, ****) & 2.35 (0.54, *) & \textbf{2.50 (0.48)} \\
    & Naturalness & 3.06 (0.58, **) & 3.21 (0.53, ns) & \textbf{3.30 (0.55)} \\
    & Informativeness & 2.44 (0.51, ****) & 2.79 (0.54, **) & \textbf{2.98 (0.53)} \\
    & Relevance & 2.34 (0.64, ****) & 2.68 (0.63, ***) & \textbf{2.89 (0.60)} \\
    \hline
    \multicolumn{2}{|c|}{Average} & 2.48 & 2.76 & \textbf{2.92} \\
    \hline
    \hline
    \multirow{4}{*}{PCSD} & Similarity & 2.01 (0.53, ****) & 2.46 (0.49, *) & \textbf{2.58 (0.43)} \\
    & Naturalness & 3.09 (0.65, ****) & 3.37 (0.54, ns) & \textbf{3.40 (0.55)} \\
    & Informativeness & 2.51 (0.59, ****) & 2.92 (0.53, *) & \textbf{3.05 (0.55)} \\
    & Relevance & 2.24 (0.69, ****) & 2.70 (0.61, **) & \textbf{2.84 (0.60)} \\
    \hline
    \multicolumn{2}{|c|}{Average} & 2.46 & 2.89 & \textbf{2.97} \\
    \hline
 \end{tabular}
 \vspace{-4mm}
\end{table}

\subsubsection{\textbf{RQ4:} Human Evaluation}
\label{subsubsec:human_evaluation}

\begin{figure}[t]
  \centering
  \includegraphics[width=\linewidth]{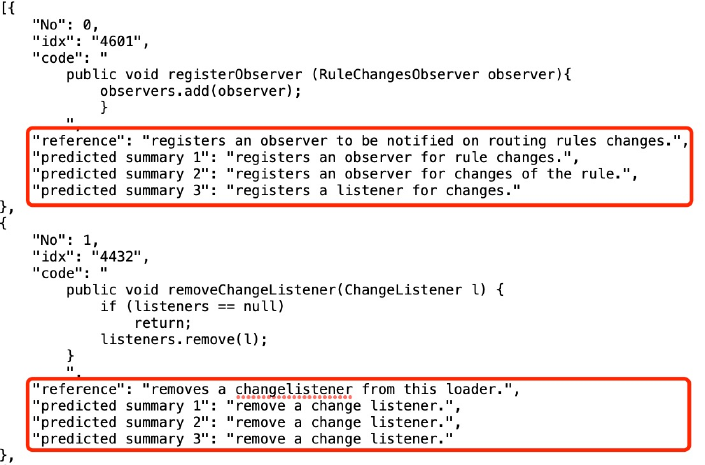}
  \vspace{-4mm}
  \caption{\revise{Example of interface of human evaluation}}
  \label{fig:human_evaluation_interface}
\end{figure}

Many works~\cite{2016-CODE-NN, 2020-Hybrid-DeepCom, 2020-R2Com, 2020-Rencos, 2021-EditSum, 2021-CAST, 2021-Reassessing-Metrics-for-Code-Summarization, 2022-Evaluation-Neural-Code-Summarization} find that it is not enough to use only automatic evaluation because the automatic metrics (BLEU, METEOR, and ROUGE-L) mainly calculate the textual similarity rather than the semantic similarity between the generated summaries and the reference summaries. Hence, we conduct a human evaluation by following the previous works~\cite{2021-SiT, 2020-Hybrid-DeepCom, 2020-R2Com, 2020-Rencos, 2021-CAST} to evaluate the summaries generated by\deletion{ the state-of-the-art baselines} SiT, UniXcoder, and our {\toolname}. \revision{The selection of SiT and UniXcoder as comparison models stems from their status as advanced code summarization techniques representing distinct paradigms: SiT is a non-pre-training-based technique, while UniXcoder is a pre-training-based technique.} Specifically, we invite ten volunteers with more than three years of programming experience and excellent English skills to carry out the evaluation. Each volunteer scores the generated summaries from 0 to 4 (the higher, the better) from four aspects: similarity (similarity of the generated summaries and the reference summaries), naturalness (grammaticality and fluency), informativeness (the amount of content carried over from the input code snippets to the generated summaries, ignoring fluency) and relevance (the degree to which the generated summaries are relevant with the input code snippets). We randomly select 100 code snippets, including 50 from the JCSD dataset and 50 from the PCSD dataset, the corresponding summaries generated by SiT, UniXcoder, and our {\toolname}, and the reference summaries (i.e., ground-truth), respectively. We divide the 100 samples into two groups, and each of them includes 50 samples, of which 25 belong to the JCSD dataset and 25 belong to the PCSD dataset. 
\revise{We place all samples to be evaluated in text files. Each participant is provided with two files containing 25 samples randomly selected from the test set of the JCSD and PCSD datasets, respectively. As shown in \deletion{Figure}\revision{Fig.}~\ref{fig:human_evaluation_interface}, for each sample, we present 7 attributes, including Number (No), the index of the sample in the test set (idx), code snippet to be summarized (code), reference summary (reference), and summaries generated by three code summarization techniques (predicted summary 1, predicted summary 2, and predicted summary 3). To prevent response bias, the specific code summarization techniques that predicted summaries 1--3 are intransparent to participants. The three techniques generating predicted summaries 1--3 do not have fixed correspondences with SiT, UniXCoder, and our {\toolname}. In other words, any summary seen by participants could be generated by any of the three techniques.} 
To reduce the workload of volunteers and ensure the fairness of experimental results, each volunteer randomly evaluates only one group of samples. Each summary is evaluated by five volunteers, and the final score is the average of them. 

The results of the human evaluation are shown in \deletion{Table}\revision{TABLE}~\ref{tab:human_evaluation}. The standard deviations of all techniques (the first values in all parentheses) are small, indicating that their scores by humans are about the same degree of concentration~\cite{2021-EditSum}. From \deletion{Table}\revision{TABLE}~\ref{tab:human_evaluation}, it can be observed that overall our {\toolname} consistently outperforms SiT and UniXcoder in all four aspects. On the JCSD dataset, compared with SiT and UniXcoder, {\toolname} improves on average by 17.88\% and 5.80\% in four aspects, respectively, while on the PCSD dataset, {\toolname} improves on average by 20.51\% and 3.67\%, respectively. 

In addition, we follow ~\cite{2021-CAST} and confirm the superiority of our {\toolname} using Wilcoxon signed-rank tests ~\cite{1963-Wilcoxon, 2014-Hitchhiker-Guide-Statistical-Tests, 2015-Practical-Test-Selection} for the human evaluation. Specifically, for each aspect, we perform the paired Wilcoxon-Mann-Whitney signed-rank test on all scores by humans for {\toolname} and each baseline (i.e., SiT or UniXcoder) at a significance level of 5\%. The test results are presented in the second values of the parentheses in \deletion{Table}\revision{TABLE}~\ref{tab:human_evaluation}. For example, ``****'' in the SiT column of the second row indicates that there is an extremely significant difference between scores by humans for {\toolname} and SiT in terms of the similarity aspect. From \deletion{Table}\revision{TABLE}~\ref{tab:human_evaluation}, it can be observed that compared with SiT and UniXcoder, the summaries generated by {\toolname} are more significantly similar to the reference summaries. For the naturalness of generated summaries, {\toolname} and UniXcoder are significantly better than SiT, which means that both {\toolname} and UniXcoder can generate more grammatically fluent summaries. For the informativeness of generated summaries, {\toolname} is better than SiT and UniXcoder, which means that {\toolname} tends to generate summaries with comprehensive semantics. For the relevance aspect, {\toolname} significantly outperforms SiT and UniXcoder, which means the summaries generated by {\toolname} are more relevant to the input code snippets.

\section{Case Study}
\label{sec:case_study}
In this section, we provide case studies to understand the generated summaries of {\toolname} compared to \deletion{the state-of-the-art }SiT\revise{, CodeBERT, }and UniXcoder\deletion{, and to demonstrate the advantages and disadvantages of our {\toolname}}.

\begin{figure*}[!t]
  \centering
  \includegraphics[width=\linewidth]{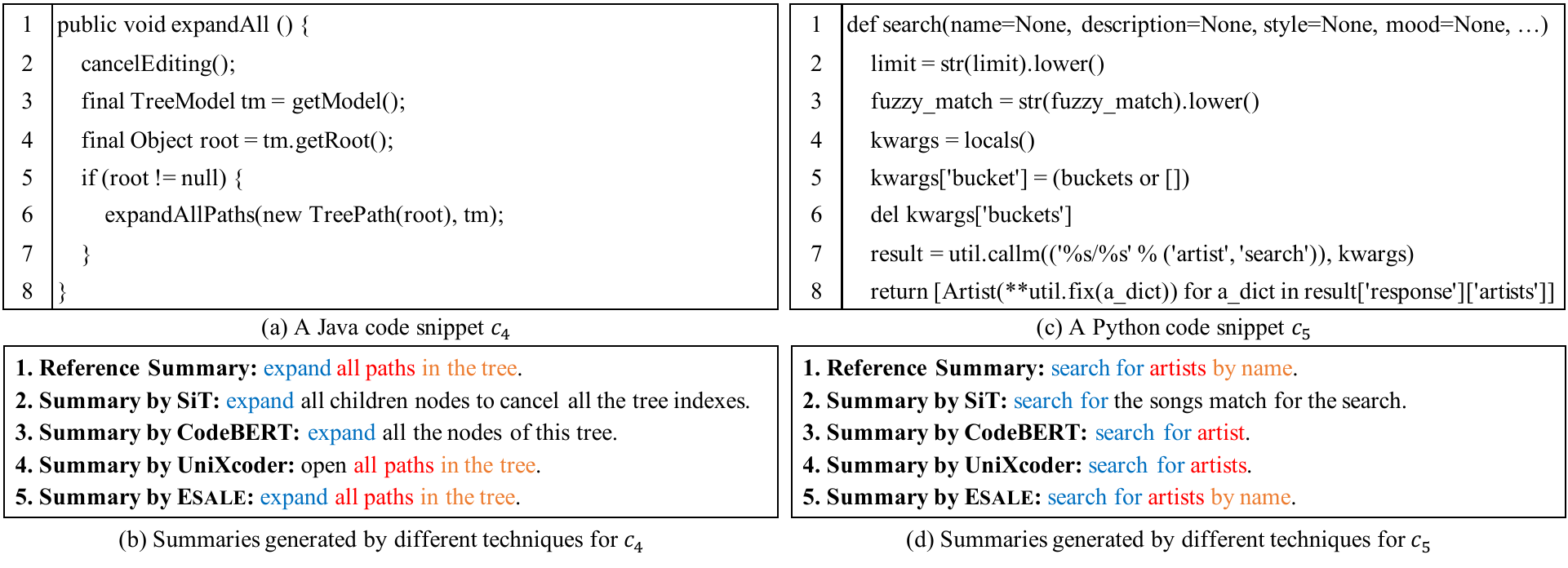}
  \vspace{-4mm}
  \caption{\delete{Two Real cases.}\revise{$c_4$ is from the test set of the JCSD with sample number 793. $c_5$ is from the test set of the PCSD with sample number 9580.}}
  \label{fig:case_1}
  \vspace{-4mm}
\end{figure*}

\deletion{Figure}\revision{Fig.}~\ref{fig:case_1} shows two real-world examples \delete{$c_5$}\revise{$c_4$} and \delete{$c_6$}\revise{$c_5$} from the test sets of the JCSD and PCSD datasets, respectively. \deletion{Figure}\revision{Fig.}~\ref{fig:case_1}(b) shows the reference summary of \delete{$c_5$}\revise{$c_4$} (line 1) and summaries generated by SiT, \revise{CodeBERT, }UniXcoder, and our {\toolname} for \delete{$c_5$}\revise{$c_4$} (lines \delete{2-4}\revise{2-5}). According to the grammar rules in natural language, we can simply divide the reference summary into three parts: ``expand'' (Blue font), ``all paths'' (Red), and ``in the tree'' (Orange font). \deletion{From this example, it}\revision{It }can be observed, compared with the reference summary, 1) both SiT\revise{ and CodeBERT} only correctly \delete{covers}\revise{cover} the first part (i.e., ``expand''); 2) UniXcoder can cover the last two parts (i.e., ``all paths" and ``in the tree"); 3) our {\toolname} can successfully cover all three parts. Similarly, for the python code snippet \delete{$c_6$}\revise{$c_5$} in \deletion{Figure}\revision{Fig.}~\ref{fig:case_1}(c), our {\toolname} can successfully cover all three parts of the reference summary, as shown in \deletion{Figure}\revision{Fig.}~\ref{fig:case_1}(d), while SiT only covers the first part (i.e., ``search for'') and\revise{ CodeBERT and} UniXcoder can cover the first two parts (i.e., ``search for" and ``artists"). In summary, based on the above two examples and observations, \deletion{we can conclude}\revision{it can be concluded} that our {\toolname} outperforms SiT\revise{, CodeBERT,} and UniXcoder in learning the mapping between code snippets and summaries, and has more a powerful code summarization performance.
Due to the page limit, please refer to our project website~\cite{2024-ESALE-google-site} for more case studies.

\section{Threats to Validity}
\label{sec:threats_to_validity}
There are three main threats to validity.

First, we cannot guarantee that the scores of human evaluation are fair. To mitigate this threat, we evaluate every generated summary by five evaluators and use the average score of the five evaluators as the final result. In addition, the standard deviations of all techniques are small (less than 0.7), indicating that scores by humans are about the same degree of concentration (detailed in Section~\ref{subsubsec:human_evaluation}).

Second, in neural network model design, there are many orthogonal aspects such as different token embeddings, whether to use beam search or teacher forcing. When showing the generality of {\toolname}, we have done the experiments in a controlled way. A future work is to do all experiments in a more controlled way, and the performance of {\toolname} could rise further when combined with all other orthogonal techniques.

Third, we use 
datasets across six programming languages to validate the effectiveness of {\toolname} we proposed in this paper. Although {\toolname} only takes token sequences of code snippets as input and does not require other complex code features (e.g., AST and CFG), we do not know whether {\toolname} is equally applicable to other programming language summarization tasks. Therefore, it is necessary to conduct experiments on more programming language datasets (e.g., C/C++ 
) to verify the reliability of {\toolname}.

\section{Related Work}
\label{sec:related_work}
Code summaries can help developers quickly understand the functionalities of the program. More and more researchers are exploring code summarization techniques, which \deletion{aims}\revision{aim} to automatically generate code summaries. 
Nowadays, \deletion{the neural machine translation (NMT)}\revision{NMT}-based models \deletion{are also}\revision{have been} widely used to generate summaries for code snippets with encoder-decoder neural networks~\cite{2016-CODE-NN, 2018-DeepCom, 2019-Ast-attendgru, 2020-Hybrid-DeepCom, 2020-Code-to-Comment-Translation, 2021-BASTS, 2021-SiT, 2021-CoCoSum, 2022-M2TS, 2022-GypSum, 2022-Assemble-Summarization-Models, 2022-SCRIPT, 2023-ICS}. For example, Iyer et al.~\cite{2016-CODE-NN} are the first to apply deep learning to automatic code summarization. They adopt LSTM networks~\cite{1997-LSTM} with attention to leverage the code vectors and generate natural language sentences that describe C\# code snippets and SQL queries. Hu et al.~\cite{2018-TL-CodeSum} use one additional encoder to encode API sequences and improve the summary generation by learning the API knowledge. Subsequently, various additional information is incorporated to further improve DL-based code summarization performance, such as abstract syntax trees (ASTs)~\cite{2018-Improving-Code-Summarization-via-DRL, 2019-Ast-attendgru, 2019-Code-Summarization-with-Extended-Tree-LSTM, 2021-BASTS, 2021-SiT, 2022-M2TS, 2022-SCRIPT},\revision{ value flows~\cite{2020-Flow2Vec}}, \revised{data flow graph~\cite{2023-SG-Trans},} code property graphs~\cite{2020-FusionGNN}, similar code snippets~\cite{2020-R2Com, 2021-EditSum}, \revised{important code statements~\cite{2023-ICS, 2024-EACS}}, file context~\cite{2020-Improved-Summarization-Attention-File-Context}, \revised{project-specific knowledge~\cite{2022-MPCos},} etc. 
\deleted{Recently}\revised{In addition}, with the success of the pre-training and fine-tuning paradigm in the field of NLP (e.g., BERT~\cite{2019-BERT} and T5~\cite{2020-T5}), many works have introduced this paradigm to further boost neural code summarization, such as CodeBERT~\cite{2020-CodeBERT}, CodeT5~\cite{2021-CodeT5}, and UniXcoder~\cite{2022-UniXcoder}. These works first pre-train a model with general language modeling tasks, such as MLM and ULM. Then, they fine-tune the pre-trained models on code summarization~\cite{2020-CodeBERT, 2021-CodeT5, 2022-UniXcoder}. However, although existing pre-trained models have achieved significant progress in general code feature learning, they are still insufficient in learning the code-summary alignment.

\revised{
Recently, with the success of large language models (LLMs) in NLP~\cite{2023-Shortcut-Learning-of-LLM-in-NLP, 2023-ChatGPT-NLP-task-solver}, an increasing number of SE researchers have started integrating them into the resolution process of various SE tasks~\cite{2023-Survey-on-LLMs-for-SE, 2023-LLMs-for-SE-Survey, 2023-Critical-Review-LLM-Program-Repair}, such as code generation~\cite{2024-Teaching-Code-LLMs-to-Repository-Level-Code-Generation}, program repair~\cite{2024-A-SLR-on-LLM-for-APR, 2024-Survey-of-Learning-based-APR}, vulnerability localization~\cite{2024-Empirical-Study-of-Vulnerability-Localization-with-LLMs}, and, of course, code summarization tasks. For example, 
Ahmed et al.~\cite{2022-Few-shot-Training-LLMs-for-Code-Summarization} investigate the effectiveness of few-shot training in adapting LLMs to code summarization and find that it can make Codex significantly outperform fine-tuned small pre-trained language models (PLMs) (e.g., CodeT5). 
Given the concern of potential code asset leakage when using commercial LLMs (e.g., GPT-3.5), Su et al.~\cite{2024-Distilled-GPT-for-Code-Summarization} utilize knowledge distillation technology to distill small models from LLMs (e.g., GPT-3.5). Their experimental findings reveal that the distilled small models can achieve comparable code summarization performance to LLMs. 
Gao et al.~\cite{2023-What-Makes-Good-In-Context-Demonstrations} investigate the optimal settings for in-context learning, including few-shot example selection methods, few-shot example order, and the number of few-shot examples. Their experimental results demonstrate that carefully designed few-shot examples can significantly improve LLMs' performance. 
Geng et al.~\cite{2024-LLM-Few-Shot-Summarizers-Multi-Intent-Comment-Generation} investigate the ability of LLMs to address multi-intent comment generation.  
Ahmed et al.~\cite{2024-Semantic-Augmentation-of-Prompts-for-Code-Summarization} propose to enhance few-shot samples with semantic facts automatically extracted from the source code. Sun et al.~\cite{2023-Automatic-Code-Summarization-via-ChatGPT} design some heuristic questions to collect the feedback of ChatGPT, thereby finding an appropriate prompt to guide ChatGPT to generate in-distribution code summaries.  
Some studies~\cite{2022-No-More-Fine-tuning-in-Code-Intelligence, 2023-CodePrompt, 2023-Prompt-CS} have also investigated the applicability of Parameter-Efficient Fine-Tuning (PEFT) techniques in code summarization tasks.  
}

\revised{
Although LLMs have been widely researched and applied due to their powerful content generation capabilities, it is important to note that there is no free lunch. 
Firstly, extensively utilizing commercial LLMs (e.g., GPT-3.5 and GPT-4) for code summarization tasks is costly. It also poses a risk of data leakage, as sensitive code might need to be transmitted to external servers. In contrast, our model can be deployed locally, ensuring that sensitive code remains within the organization's secure environment.
Secondly, deploying open-source LLMs independently is also expensive for users or organizations due to their significant computational requirements. The more advanced LLMs typically have more parameters, and to ensure these models can efficiently generate useful outputs, extensive and costly hardware (such as GPUs) is essential. Our model offers a cost-effective alternative that can be deployed with less expensive hardware.
Furthermore, most LLMs are designed to be general-purpose (to support multiple downstream tasks) and may not perform optimally on specific downstream tasks without extensive fine-tuning. Despite the advancements in prompt engineering techniques, the performance of LLMs on code summarization tasks may not significantly surpass that of smaller specialized models trained using supervised learning. For example, the work by Ahmed et al.~\cite{2022-Few-shot-Training-LLMs-for-Code-Summarization} demonstrates that the performance of the adapted LLM Codex on Python and PHP code summarization tasks is not significantly better than that of the smaller model CodeT5~\cite{2021-CodeT5}. Our model, however, is optimized specifically for code summarization, ensuring more reliable and accurate performance for this task. 
Finally, our small model allows for greater flexibility and customizability for specific scenario/domain requirements. This adaptability is particularly beneficial for users or organizations with unique needs that may not be fully met by general-purpose LLMs. 
}

\section{Conclusion}
\label{sec:conclusion}
In this paper, we propose an approach for code summarization, namely {\toolname}, which improves the code summarization performance by enhancing the encoder to code-summary alignment. {\toolname} is first trained using \revise{a} multi-task learning paradigm with \deletion{two general tasks and one domain-specific task}\revision{three summary-focused tasks}, and then fine-tuned on the code summarization task. We conduct quantitative comprehensive experiments and qualitative human evaluations to verify the effectiveness of {\toolname}. And all results show that our {\toolname} is significantly better than state-of-the-art baselines.

\ifCLASSOPTIONcompsoc
  \section*{Acknowledgments}
\else
  \section*{Acknowledgment}
\fi

The authors would like to thank the anonymous reviewers for their insightful comments. The authors thank Guanhong Tao (taog@purdue.edu) for his constructive comments and discussions on the manuscript. This work is supported partially by the National Natural Science Foundation of China (61932012, 62372228), and the National Research Foundation, Singapore, and the Cyber Security Agency under its National Cybersecurity R\&D Programme (NCRP25-P04-TAICeN). Any opinions, findings and conclusions or recommendations expressed in this material are those of the author(s) and do not reflect the views of National Research Foundation, Singapore and Cyber Security Agency of Singapore.

\ifCLASSOPTIONcaptionsoff
  \newpage
\fi

\bibliographystyle{IEEEtran}
\bibliography{reference}


\end{document}